\documentclass[preprint,review,12pt]{elsarticle}

\usepackage{graphicx}  
\usepackage{dcolumn}  
\usepackage{bm}           
\usepackage{amsmath}
\usepackage{mathrsfs}
\usepackage{float}
\usepackage{listings}
\usepackage{multirow}
\usepackage{subfigure}
\usepackage{amsfonts}
\usepackage{lineno} 

\newcommand{\vx}{\mathbf{x}}

\makeatletter
\def\ps@pprintTitle{%
 \let\@oddhead\@empty
 \let\@evenhead\@empty
 \def\@oddfoot{\centerline{\thepage}}%
 \let\@evenfoot\@oddfoot}
\makeatother 

\begin{document}

\begin{frontmatter}

%
%

\title{Anomalous transport in disordered fracture networks: spatial Markov model for dispersion with variable injection modes}

%
%

\author{Peter K. Kang}
\address{Korea Institute of Science and Technology, Seoul 02792, Republic of Korea}
\address{Massachusetts Institute of Technology, 77
  Massachusetts Ave, Building 1, Cambridge, Massachusetts 02139, USA}
  
\author{Marco Dentz}
\address{Institute of Environmental Assessment and Water Research
  (ID\mbox{\AE}A), Spanish National Research Council (CSIC),
08034 Barcelona, Spain}

\author{Tanguy Le Borgne}
\address{Universit\'e de Rennes 1, CNRS, Geosciences Rennes,
  UMR 6118, Rennes, France}

\author{Seunghak Lee}
\address{Korea Institute of Science and Technology, Seoul 02792, Republic of Korea}

\author{Ruben Juanes}
\address{Massachusetts Institute of Technology, 77
  Massachusetts Ave, Building 1, Cambridge, Massachusetts 02139, USA}
  
\date{\today}
\begin{abstract}
	We investigate tracer transport on random discrete fracture networks that are characterized by the statistics of the fracture geometry and hydraulic conductivity. While it is well known that tracer transport through fractured media can be anomalous and particle injection modes can have major impact on dispersion, the incorporation of injection modes into effective transport modelling has remained an open issue. The fundamental reason behind this challenge is that---even if the Eulerian fluid velocity is steady---the Lagrangian velocity distribution experienced by tracer particles evolves with time from its initial distribution, which is dictated by the injection mode, to a stationary velocity distribution. We quantify this evolution by a Markov model for particle velocities that are equidistantly sampled along trajectories.  
This stochastic approach allows for the systematic incorporation of the initial velocity distribution and quantifies the interplay between 
velocity distribution and spatial and temporal correlation. The proposed spatial Markov model is characterized by the initial velocity distribution, which is determined by the particle injection mode, the stationary Lagrangian velocity distribution, which is derived from the Eulerian velocity distribution, and the spatial velocity correlation length, which is related to the characteristic fracture length. 
This effective model leads to a time-domain random walk for the evolution of particle positions and velocities, whose joint distribution follows a Boltzmann equation.  Finally, we demonstrate that the proposed model can successfully predict anomalous transport through discrete fracture networks with different levels of heterogeneity and arbitrary tracer injection modes.
\end{abstract}

\end{frontmatter}

\section{Introduction}
Flow and transport in fractured geologic media control many important natural and engineered processes, including nuclear waste disposal, geologic carbon sequestration, groundwater contamination, managed aquifer recharge, and geothermal production in fractured geologic media~\citep[e.g.,][]{bodvarsson99, lewicki2007natural, tang1981contaminant, chrysikopoulos2009removal, pruess06-egs}. Two dominant approaches exist for simulating flow and transport through fractured media: the equivalent porous medium approach~\citep{neumanwinter87, tsangtsang96} and the discrete fracture network approach (DFN)~\citep{kiraly79, cacas1990DFN, nordqvist1992variable, morenoneretnieks93, juanes02-ijnme, park2003transport, karimidurlofsky04, martinezlanda05, frampton2011numerical, dreuzy2012influence, schmidgeiger13, makedonska2015particle, hyman2015dfnworks, Bernabe16}. The DFN approach explicitly resolves individual fractures whereas the equivalent porous medium approach represents the fractured medium as a single continuum by deriving effective parameters to include the effect of the fractures on the flow and transport. The latter, however, is hampered by the fact that a representative elementary volume may not exist for fractured media~\citep{bear72, marsily-qh}. Dual-porosity models are in between these two approaches, and conceptualize the fractured-porous medium as two overlapping continua, which interact via an exchange term~\citep{barenblatt60-fissured, warrenroot63, kazemi76, bibby81-dual, feenstra85, maloszewskizuber85, pruess85-minc, arbogast1990dual, gerke1993dual}. 

DFN modelling has advanced significantly in recent years with the increase in computational power. Current DFN simulators can take into account multiple physical mechanisms occurring in complex 3D fracture systems. Recent studies also have developed methods to explicitly model advection and diffusion through both the discrete fractures and the permeable rock matrix~\citep{geiger2010, houseworth2013, willmann2013, sebben2016}. In practice, however, their application must account for the uncertainty in the subsurface characterization of fractured media, which is still an considerable challenge~\citep{chenhubbard06, dorn12-wrr, kang16-flowseismic}. Thus, there is a continued interest in the development of upscaled transport models that can be parameterized with a small number of model parameters. Ideally, these model parameters should have a clear physical interpretation and should be determined by means of field experiments, with the expectation that the model can then be used for predictive purposes~\citep{beckershapiro03, kang15-wrr}.

Developing an upscaled model for transport in fractured media is especially challenging due to the emergence of anomalous (non-Fickian) transport. While particle spreading is often described using a Fickian framework, anomalous transport---characterized by scale-dependent spreading, early arrivals, long tails, and nonlinear scaling with time of the centered mean square displacement---has been widely observed in porous and fractured media across multiple scales, from pore~\citep{seymour2004anomalous, scheven05, bijeljic11-prl, kangdeanna14, gjetvaj2015dual} to single fracture~\citep{detwilerrajaram00, drazerauradou04, wang2014, kang2016emergence} to column~\citep{hatano1998, cortis2004} to field scale~\citep{garabedian91, becker2000, haggerty01-2, haggerty01-3, leborgnegouze08, kang15-wrr, hyman2015influence}. The ability to predict anomalous transport is essential because it leads to fundamentally different behavior compared with Fickian transport~\citep{shlesinger74, bouchaudgeorges90, metzlerklafter00}. 

The continuous time random walk (CTRW) formalism~\citep{schermontroll75, klaftersilbey80} is a framework to describe anomalous transport through which models particle motion through a random walk in space and time characterized by random space and time increments, which accounts for variable mass transfer rates due to spatial heterogeneity. It has been used to model transport in heterogeneous porous and fractured media~\citep{berkowitzscher97-prl, berkowitzcortis06, dentzcortis04, geigercortis10, kangdentz11-pre, wang2014, dentz2015} and allows  incorporating information on flow heterogeneity and medium geometry for large scale transport modelling. Similarly, the time-domain random walk  (TDRW) approach~\cite[][]{benkepainter03, paintercvetkovic05,  paintertdrw2008} models particle motion due to distributed space and time increments, which are derived from particle velocities and their correlations. 
The analysis of particle motion in heterogeneous flow fields demonstrate that Lagrangian particle velocities
exhibit sustained correlation along their trajectory~\citep{cvetkovic1996, benkepainter03, leborgnedentz08-prl, gotovac2009,
  meyertchelepi10, kangdentz11-prl, deanna13-prl, datta13-prl,
  kangdeanna14}. Volume conservation induces correlation in the Eulerian
velocity field because fluxes must satisfy the divergence-free
constraint. This, in turn, induces correlation in
the Lagrangian velocity along a particle trajectory. 
To take into account velocity correlation, Lagrangian models based on Markovian processes have been
proposed~\citep{benkepainter03, paintercvetkovic05, leborgnedentz08-prl, meyertchelepi10, kangdentz11-prl, deanna13-prl,  kangdeanna14, kang15-wrr, kang15-pre}. 
Spatial Markov models are based on the observation that successive velocity transitions measured equidistantly along the mean flow direction exhibit Markovianity: a particle's velocity at the next step is fully determined by its current velocity. The spatial Markov model, which accounts for velocity correlation by incorporating this one-step velocity correlation information, has not yet been extended to disordered (unstructured) DFNs. 

The mode of particle injection can have a major impact on transport through porous and fractured media~\citep{kreft1978, sposito1994, leborgne07, frampton2009significance, jankovic2010analysis, leborgnedentz10-awr, hyman2015influence, dagan2016solute}. Two generic injection modes are uniform (resident) injection and flux-weighted injection with distinctive physical meanings as discussed in \citet{frampton2009significance}. The work by \citet{sposito1994} is one of the earliest studies of the impact of different particle injection modes on the time evolution of a solute plume spatial moments. The significance of injection modes on particle transport through discrete fracture networks has been studied for fractured media \citep{frampton2009significance, hyman2015influence}. \citet{dagan2016solute} recently clarified the theoretical relation between injection modes and plume mean velocity. Despite recent advances regarding the significance of particle injection modes, the incorporation of injection methods into effective transport modelling is still an open issue \citep{frampton2009significance}. The fundamental challenge is that the Lagrangian velocity distribution experienced by tracer particles evolves with time from its initial distribution which is dictated by the injection mode to a stationary velocity distribution \citep{cvetkovic1996, frampton2009significance, gotovac2009, DentzVCTRW2016}. In this paper, we address these fundamental questions, in the context of anomalous transport through disordered DFNs.

The paper proceeds as follows. In the next section, we present the studied
random discrete fracture networks,
the flow and transport equations and details of the different particle injection rules. In Section~\ref{Sec:3}, we investigate the emergence of anomalous transport by direct
Monte Carlo simulations of flow and particle
transport. In Section~\ref{Sec:4}, we analyze 
Eulerian and Lagrangian velocity statistics to gain insight into the effective
particle dynamics and elucidate the key mechanisms that lead to the observed anomalous
behavior. In Section~\ref{Sec:5}, we develop a spatial Markov model that is
characterized by the initial velocity distribution, probability density function (PDF) of Lagrangian
velocities and their transition PDF, which are derived from the
Monte Carlo simulations. The proposed model is in  
excellent agreement with direct Monte Carlo simulations. We then present a parsimonious spatial Markov model that quantifies velocity correlation with a single parameter. The predictive
capabilities of this simplified model are demonstrated by comparison to the
direct Monte Carlo simulations with arbitrary injection modes. In Section~\ref{Sec:6}, we summarize the main findings and conclusions. 

\section{Flow and Transport in Discrete Random Fracture Networks}
\subsection{Random Fracture Networks}
\begin{figure}
  \centering
	  \includegraphics[width=3.5in]{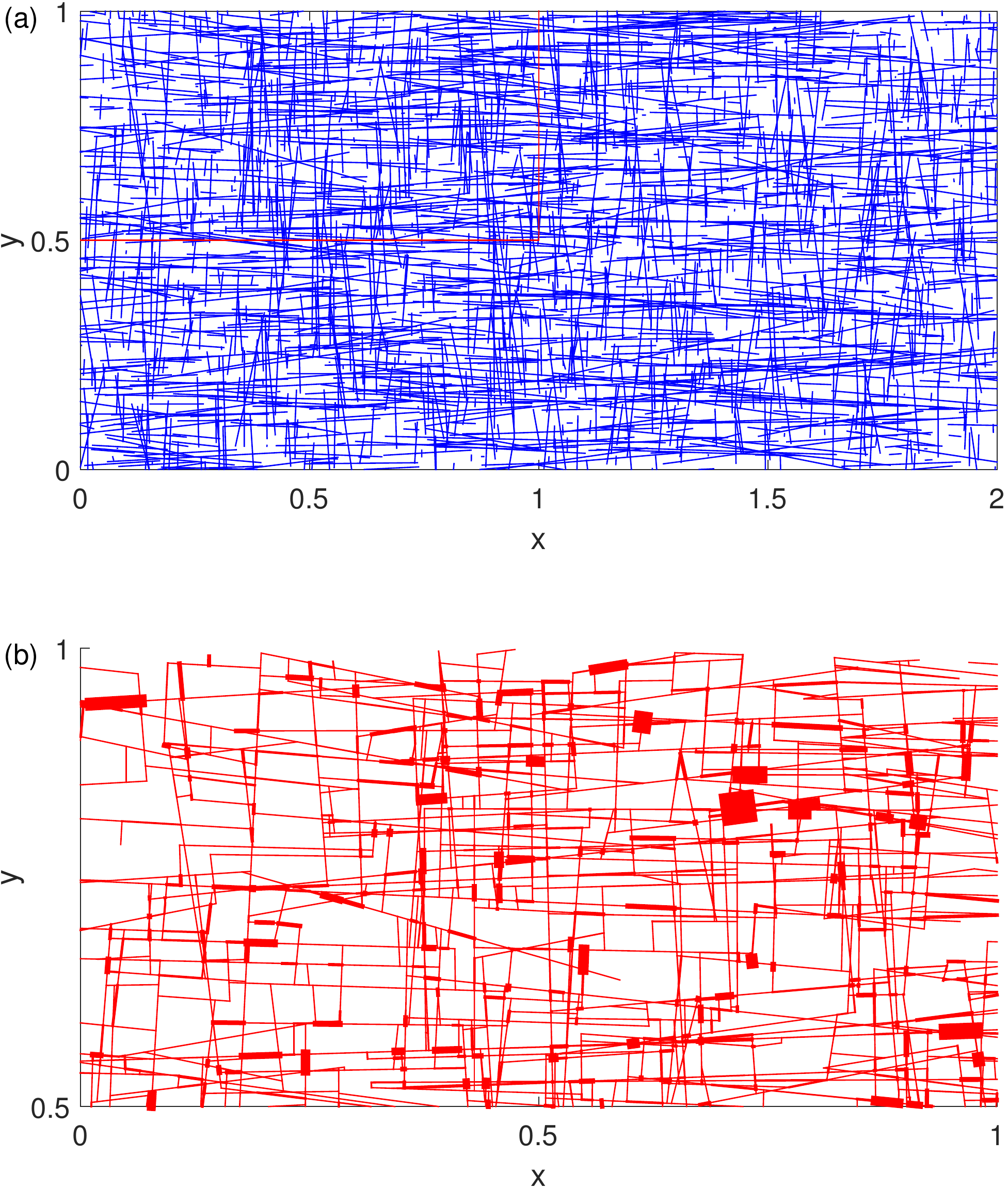}	  	  
  \caption{(a)~Example of a two-dimensional DFN studied here, with $2000$ fractures ($1000$ fractures for each fracture set). (b)~Subsection of a spatially uncorrelated conductivity field between $0 \leq x \leq 1$ and $0.5 \leq y \leq 1$. Conductivity values are assigned from a lognormal distribution with $\sigma_{\text{ln}K} = 1$. Link width is proportional to the conductivity value; only connected links are shown.}\label{fig:DFN}
\end{figure}

We numerically generate random DFNs in two-dimensional rectangular regions, and solve for flow and tracer transport within these networks. The fracture networks are composed of linear fractures embedded in an impermeable rock matrix. The idealized 2D DFN realizations are generated by superimposing two different sets of fractures, which leads to realistic discrete fracture networks~\citep{long1982, geigercortis10}.
Fracture locations, orientations, lengths and hydraulic conductivities are generated from predefined distributions, which are assumed to be statistically independent: (1)~Fracture midpoints are selected randomly over the domain size of $L_x \times L_y$ where $L_x =2$ and $L_y = 1$; (2)~Fracture orientations for two fracture sets are selected randomly from Gaussian distributions, with means and standard deviation of $0^{\circ}\pm 5^{\circ}$ for the first set, and $90^{\circ} \pm 5^{\circ}$ for the second set; (3)~Fracture lengths are chosen randomly from exponential distributions with mean $L_x/10$ for the horizontal fracture set and mean $L_y/10$ for the vertical fracture set; (4)~Fracture conductivities are assigned randomly from a predefined log-normal distribution. An example of a random discrete fracture network with $2000$ fractures is shown in Figure~\ref{fig:DFN}. 

The position vector of node~$i$ in the fracture network is denoted by $\vx_i$. The link length between nodes $i$ and $j$ is 
denoted by $l_{ij}$. The network is characterized by the distribution of link lengths 
$p_l(l)$ and hydraulic conductivity $K$. The PDF of link lengths here is 
exponential 
\begin{align}
p_l(l) = \frac{\exp(-l/\bar l)}{\bar l}. 
\end{align}
Note that the link length and orientation are independent. The characteristic fracture link length is obtained by taking the average of a link length over all the realizations, which gives $\bar{l} \approx L_x/200$.
A realization of the random discrete fracture network is generated by assigning
independent and identically distributed random hydraulic conductivities $K_{ij} > 0$ 
to each link between nodes~$i$ and~$j$.  Therefore, the $K_{ij}$ values in different links are uncorrelated. 
The set of all realizations of the spatially random network generated
in this way forms a statistical ensemble that is stationary and
ergodic. We assign a lognormal distribution of $K$ values, and study the impact
of conductivity heterogeneity on transport by varying the
variance of $\ln(K)$. We study log-normal conductivity distributions with four different variances: $\sigma_{\text{ln}K} = 1, 2, 3, 5$. The use of this particular distribution is motivated
by the fact that conductivity values in many natural media can be
described by a lognormal law~\citep{bianchi1969, vilaguadagnini06}.

\subsection{Flow Field}
Steady state flow through the network is modeled by Darcy's law~\citep{bear72} 
for the fluid flux $u_{ij}$ between nodes $i$ and $j$,  
$u_{ij}=-K_{ij}(\Phi_j-\Phi_i)/l_{ij}$, where $\Phi_i$ and $\Phi_j$ are the
hydraulic heads at nodes $i$ and $j$.
Imposing flux conservation at each node~$i$,
$\sum_j u_{ij}=0$ (the summation is over nearest-neighbor nodes),
leads to a linear system of equations, which is solved for the 
hydraulic heads at the nodes. The fluid flux through a link from node $i$ to
$j$ is termed incoming for node~$i$ if $u_{ij} < 0$, and outgoing if $u_{ij} >
0$. We denote by $\mathbf e_{ij}$ the unit vector in the direction of
the link connecting nodes $i$ and $j$.

\begin{figure}
  \centering
	  \includegraphics[width=5.5in]{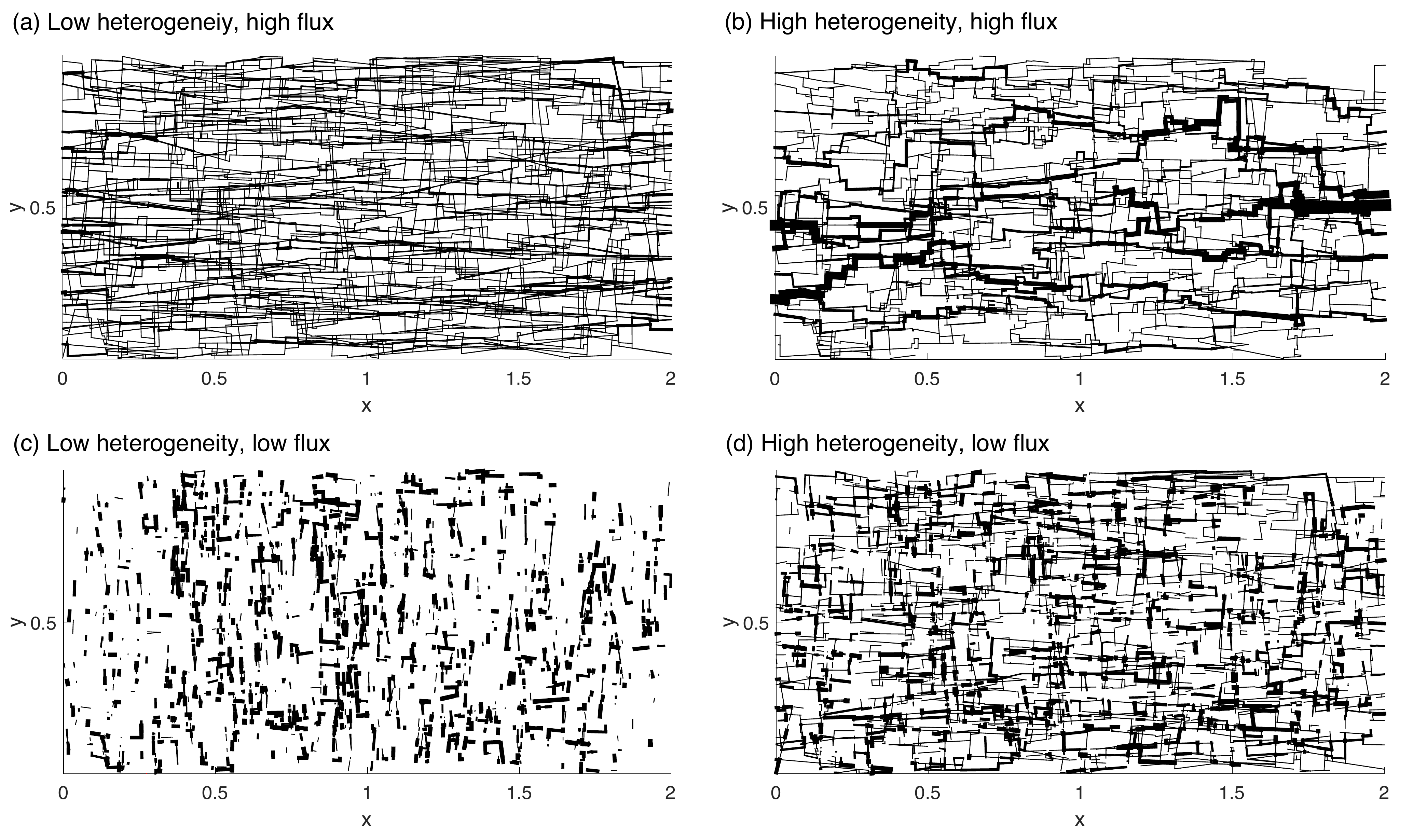}
  \caption{Normalized flow field ($|u_{ij}|/ \bar{u}$) showing high and low flux zones for a log-normal conductivity distribution with two different heterogeneities. Link width is proportional to the magnitude of the normalized flow. (a)~$\sigma_{\text{ln}K} = 1$. Links with the flux value \emph{smaller} than $ \bar{u} / 5$ are removed. (b)~$\sigma_{\text{ln}K} = 5$. Links with the flux value \emph{smaller} than $ \bar{u} / 5$ is removed. Preferential flow paths emerge as conductivity heterogeneity increases. (c)~$\sigma_{\text{ln}K} = 1$. Links with the flux value \emph{larger} than $ \bar{u} / 5$ are removed. Most of low flux values occur at the links perpendicular to the mean flow direction. (d)~$\sigma_{\text{ln}K} = 5$. Links with the flux value \emph{larger} than $ \bar{u} / 5$ are removed. Low flux values show less spatial correlation than high flux values.}\label{fig:flowfield}
\end{figure}

\begin{figure}
  \centering
	  \includegraphics[width=3.5in]{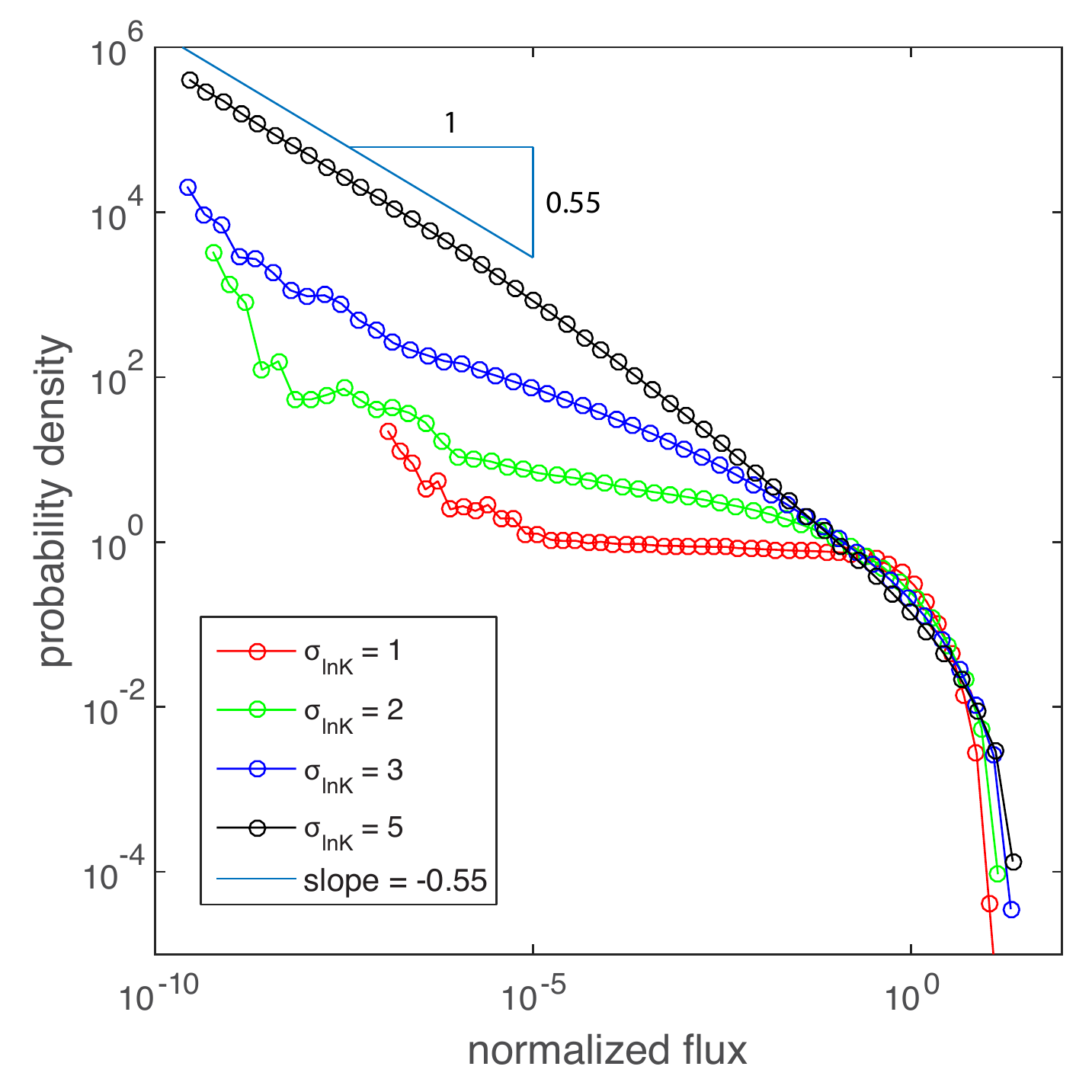}
  \caption{Eulerian flux probability density functions for four different levels of conductivity heterogeneity. Increase in conductivity heterogeneity significantly increases the probability of small flux values.}\label{fig:EulerianVpdf}
\end{figure}

We study a uniform flow setting characterized by constant mean flow in the positive
$x$-direction parallel to the principal set of factures. No-flow conditions are imposed at the top and
bottom boundaries of the domain, and fixed hydraulic head at
the left ($\Phi=1$) and right ($\Phi=0$) boundaries. The overbar in the
following denotes the ensemble average over all network realizations.  
The one-point statistics of the flow field are characterized by the Eulerian 
velocity PDF, which is obtained by spatial and ensemble sampling of  the  velocity magnitudes in the network
%
%
\begin{align}
\label{pe}
p_e(u) = \frac{\sum_{i > j} \overline{ l_{ij} {\delta(u - u_{ij})}}}{N_\ell \bar l}.  
\end{align}
where $N_\ell$ is the number of links in the network. 
Link length and flow velocities here are independent. Thus, the Eulerian velocity PDF is given by 
\begin{align}
\label{p:eulerian}
p_e(u) = \frac{1}{N_\ell} \sum_{i > j} \overline{\delta(u - u_{ij})}.  
\end{align}
Even though the underlying conductivity field is uncorrelated, the mass conservation
constraint together with heterogeneity leads to the formation of
preferential flow paths with increasing network heterogeneity~\citep{Bernabe98, kang15-pre}. This is illustrated in Figures~\ref{fig:flowfield}a and~b, which show maps of the relative velocity magnitude for high velocities in networks with log-$K$ variances of $1$ and $5$. As shown in Figures~\ref{fig:flowfield}c and~d, for low heterogeneity most small flux values occur along links perpendicular to the mean flow direction, whereas low flux values do not show directionality for the high heterogeneity case. This indicates that fracture geometry dominates small flux values for low heterogeneity and fracture conductivity dominates small flux values for high heterogeneity. An increase in conductivity heterogeneity leads to a broader Eulerian velocity PDF,
with significantly larger probability of having small flux values as illustrated in Figure~\ref{fig:EulerianVpdf}, which 
shows $p_e(u)$ for networks of different heterogeneity strength.

\subsection{Transport}
Once the fluxes at the links have been determined, we simulate transport of a
passive tracer by particle tracking. Particles are injected along a line at the inlet, $x=0$, with two different injection methods: (1)~uniform injection, and (2)~flux-weighted injection. \emph{Uniform (resident) injection} introduces particles uniformly throughout the left boundary; this means that an equal number of particles is injected into each inlet node $i_0$, 
\begin{align}
\label{random}
N_{i_0} = \frac{N_p}{\sum_{i_0}}, 
\end{align}
where $N_{i_0}$ is the number of particles injected at node $i_0$, $N_{p}$ is the 
total number of injected particles. \emph{Flux-weighted injection} introduces particles proportional to the total incoming flux $Q_{i_0}$ at the injection location $i_0$
\begin{align}
\label{flux}
N_{i_0} = N_p \frac{Q_{i_0}}{\sum_{i_0} Q_{i_0}}. 
\end{align}
Uniform injection simulates an initial distribution of tracer particles extended uniformly over a region much larger than the characteristic heterogeneity scale, and flux-weighted injection simulates a constant concentration pulse where the injected mass is proportional to the local injection flux at an inlet boundary that is much larger than the heterogeneity scale. For the uniform injection, the initial velocity distribution is then equal to the distribution of the Eulerian velocities. For the flux-weighted injection, the initial velocity distribution is equal to the flux-weighted Eulerian distribution. In general the initial velocity distribution may be arbitrary and depends on the conditions at the injection location. More detailed discussions can be found in section~\ref{Sec:4} and section~\ref{Sec:5}.

Injected particles are advected with the flow velocity $u_{ij}$ between nodes. To focus on the impact of conductivity
    variability on particle transport, we assume porosity to be constant. 
    This is a reasonable assumption because
    the variability in porosity is significantly smaller than the
    the variability in conductivity~\citep{bear72, freeze1977}. 
    
At the nodes, we apply a complete mixing rule~\citep{berkowitz1994, stockman1997, park2001}. Complete
mixing assumes that P\'eclet numbers at the nodes are small enough that
particles are well mixed within the node. Thus, the link through which
the particle exits a node is chosen randomly with
flux-weighted probability. A different node-mixing rule, streamline routing, assumes that P\'eclet numbers at nodes are large enough that particles essentially follow the streamlines and do not transition between streamlines. The complete
mixing and streamline routing rules are two end members. The local
P\'eclet number and the intersection geometry determine the
strength of mixing at nodes, which is in general between these two end-members. The impact of the mixing rule on transverse spreading can be significant for regular DFNs with low heterogeneity~\citep{park2001effects, kang15-pre}. However, its impact 
is much more limited for random DFNs~\citep{park2001}. Since our interest in this study is the longitudinal spreading in random DFNs, we 
focus on the case of complete mixing. Thus, the particle transition probabilities $p_{ij}$ from
node $i$ to node $j$ are given by
\begin{equation}
\label{eq:completemixing}
p_{ij} = \frac{|u_{ij}|}{\sum_{k} |u_{ik}|},
\end{equation}
where the summation is over outgoing links only, and $p_{ij} = 0$ for
incoming links. Particle transitions are determined only by the outgoing flux
distribution. 

\begin{figure}
  \centering
	  \includegraphics[width=5in]{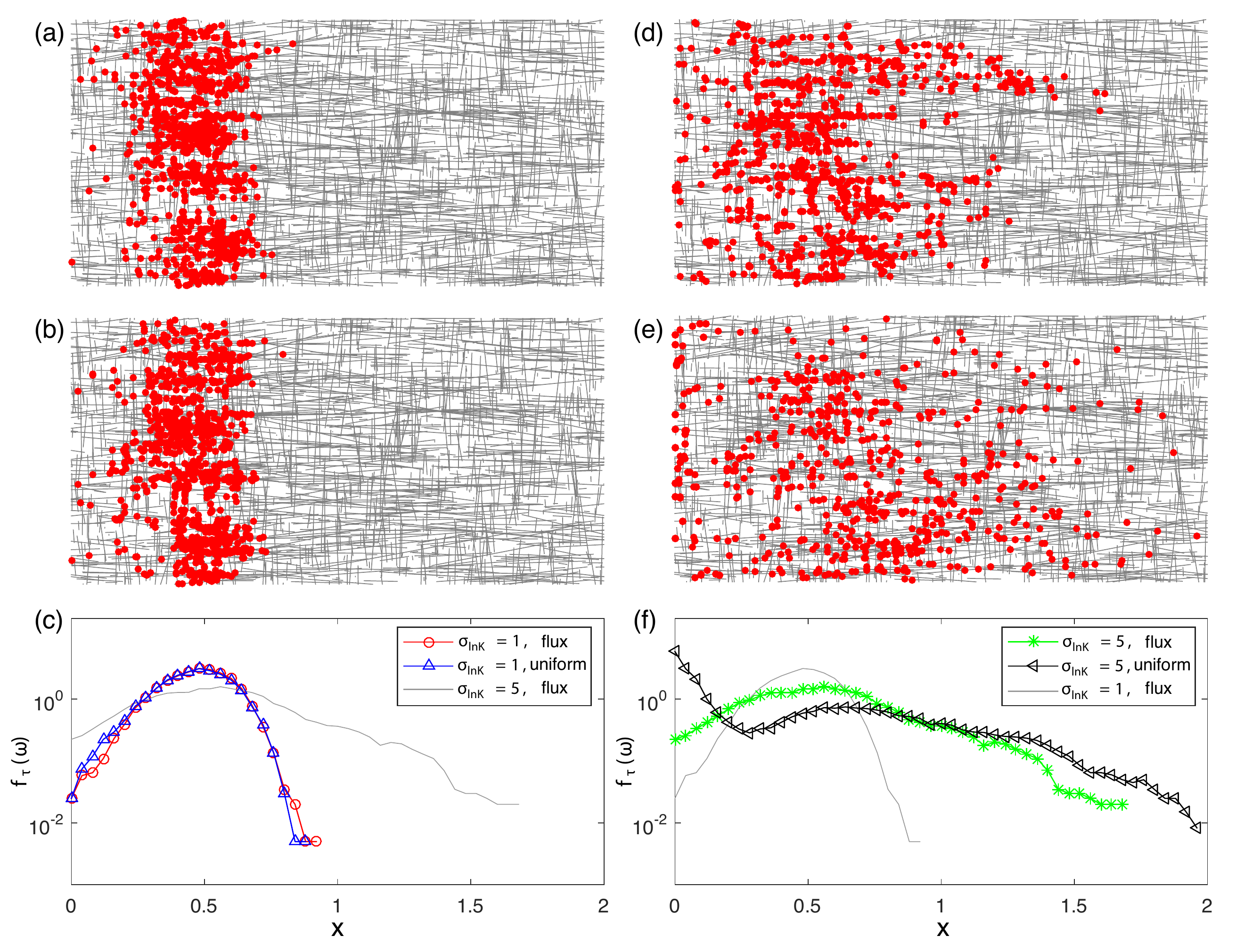}
  \caption{Particle distribution at $t = 20
    \overline{t_l}$ for a given realization after the instantaneous release of particles at the inlet, $x=0$. $\overline{t_l}$ is the median transition time to travel $L_x/100$. (a)~The low heterogeneity case ($\sigma_{\text{ln}K} = 1$) with flux-weighted injection. (b)~The low heterogeneity case ($\sigma_{\text{ln}K} = 1$) with uniform injection. (c)~The projected particle distribution in the longitudinal direction for the low heterogeneity case ($\sigma_{\text{ln}K} = 1$). (d)~The high heterogeneity case ($\sigma_{\text{ln}K} = 5$) with the flux-weighted injection. (e)~The high heterogeneity ($\sigma_{\text{ln}K} = 5$) with the uniform injection. (f)~The projected particle distribution in the longitudinal direction for the high heterogeneity case ($\sigma_{\text{ln}K} = 5$). For the high heterogeneity case, the injection method has significant impact on particle spreading. The uniform injection method leads to more anomalous spreading.}\label{fig:mixing_heterogeneity}
\end{figure}

The particle pathways and times are obtained by the following recursion relations
\begin{subequations}
\label{eq:langevin}
\begin{align}
\vx_{n+1} &= \vx_{n} + \ell_n \mathbf e_n, 
\\
t_{n+1} &= t_n + \frac{\ell_n}{u_n}, 
\end{align}
\end{subequations}
 where $\vx_n \equiv \vx_{i_n}$ is the particle position after $n$ random walk steps, $\ell_n \equiv l_{i_n i_{n+1}}$ the 
particle displacement and $\mathbf e_n \equiv \mathbf e_{i_ni_{n+1}}$ its orientation; the particle velocity at the $n$th step 
is denoted by $u_n \equiv |u_{i_ni_{n+1}}|$. The particle displacement, orientation and velocity determined by the transition probability $p_{i_nj}$ from node $i_n$ to the neighboring nodes $j$ given by Eq.~\eqref{eq:completemixing}. 
Equations~\eqref{eq:langevin} describe coarse-grained particle
transport for a single realization of the spatial random
network. Particle velocities and thus transition times depend on the
particle position. The particle position at time~$t$
is~$\vx(t) = \vx_{i_{n_t}}$, where 
\begin{align}
\label{nt}
n_t = \sup(n|t_n \leq t)
\end{align}
denotes the number of steps needed to reach
time~$t$. We solve transport in a single disorder realization by particle tracking based on Eq.~\eqref{eq:langevin} with the two different injection rules~\eqref{random} and~\eqref{flux} at the inlet at $x=0$. 
The particle density in a single realization is 
\begin{align}
p(\vx,t) = \langle\delta(\vx-\vx_{{n_t}})\rangle,
\end{align}
where the angular brackets denote the average over all injected particles. 
As shown in Figure~\ref{fig:mixing_heterogeneity}, both network heterogeneity and
injection rule have significant impact on particle
spreading. An increase in network heterogeneity leads to an increase in longitudinal particle spreading, and the uniform injection rule significantly enhances longitudinal spreading compared to flux-weighted injection. The impact of network heterogeneity and injection method can be clearly seen from projected concentration profiles, $f_{\tau}(\omega)$. Arbitrary injection modes are also studied and discussed in section~\ref{Sec:5.2}.

%

\section{Average Solute Spreading Behavior\label{Sec:3}}

We first study the average solute spreading behavior for the
four different levels of conductivity heterogeneity and the two different injection methods described above. 
We first illustrate the persistent effect of the particle injection method on particle transport, with the two different injection modes. 
To investigate the average spreading behavior, we average over all particles and 
network realizations. The average particle density is given by 
\begin{align}
\label{eq:averageP}
\overline{ P}(\vx,t)= \overline{ \langle\delta(\vx-\vx_{n_t})\rangle },
\end{align}
where the overbar denotes the ensemble average over all realizations.
We run Monte Carlo particle tracking simulations for $100$
realizations for each combination of conductivity heterogeneity and particle injection
rule. In each realization, we release $10^4$ particles at the inlet ($x=0$) with the two different injection methods. 
\subsection{Breakthrough Curves\label{sec:btc}}
The average particle spreading behavior is first studied with the first passage time distribution (FPTD) or breakthrough curve (BTC) of particles at a control plane located at $x=x_c$. The FPTD is obtained by averaging over the individual particle arrival times $\tau_a(x_c) = \text{inf}(t_n|\;|x_n - x_0| > x_c)$ as 

\begin{align}
\label{f:fpt}
f(\tau,x_c) = \overline{\langle \delta[\tau - \tau_a(x_c)] \rangle}. 
\end{align}

\begin{figure}
  \centering
	  \includegraphics[width=5in]{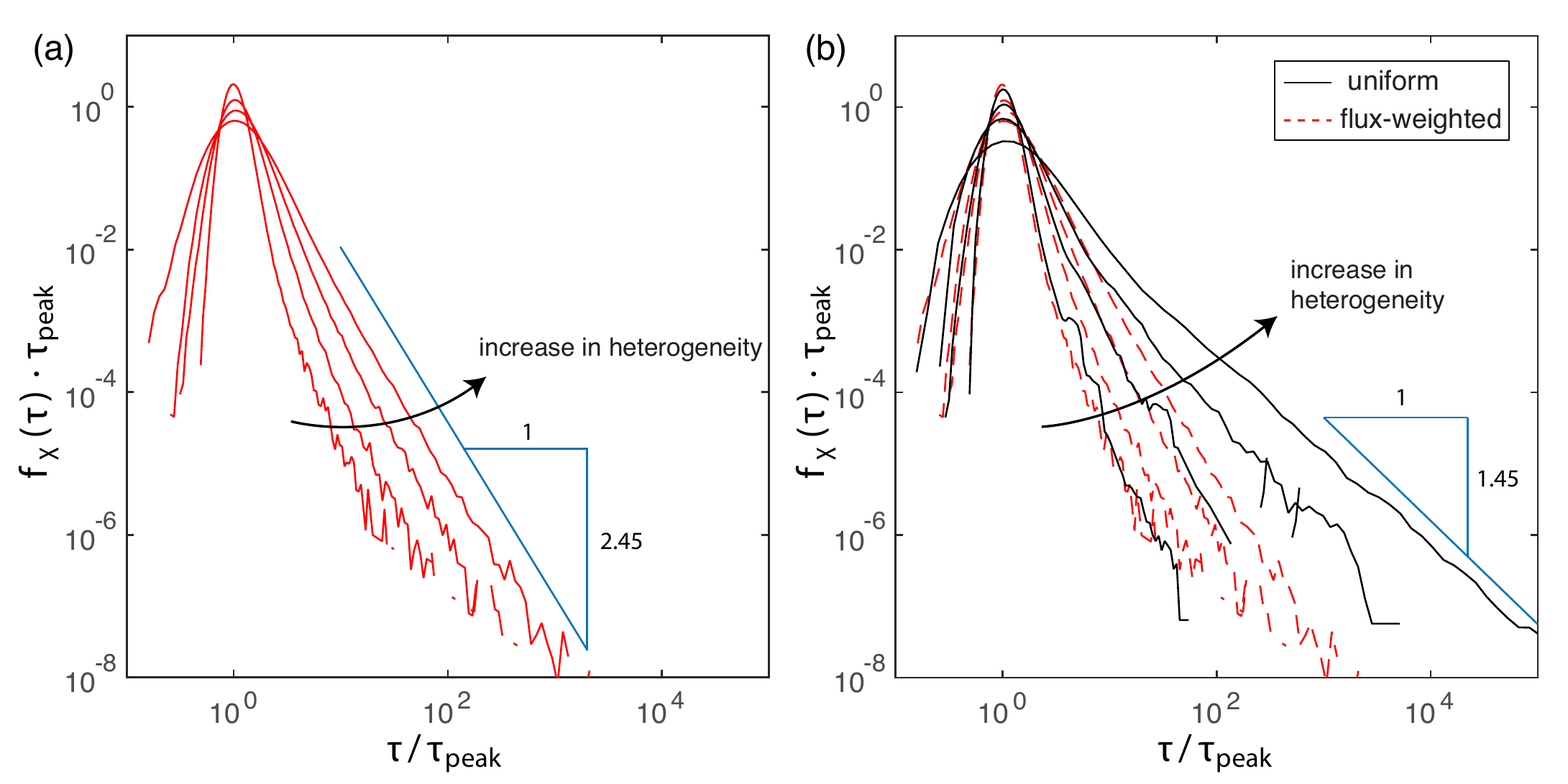}
  \caption{(a) FPTDs for $\sigma_{\ln K} = 1, 2, 3, 5$ with flux-weighted injection at $x_c = 200 \bar l$. Increase in conductivity heterogeneity leads to larger dispersion and stronger late-time tailing. (b) FPTDs for $\sigma_{\ln K} = 1, 2, 3, 5$ with uniform injection (black solid lines). Uniform injection leads to significantly larger dispersion and late-time tailing compared to the flux-weighted injection (red dashed lines). FPTDs are normalized with the peak arrival time.
}\label{fig:BTC_MC}
\end{figure}

 Figure~\ref{fig:BTC_MC} shows FPTDs at the outlet, $f(\tau,x_c= 200 \bar l)$, for different conductivity heterogeneities and injection
rules. Conductivity heterogeneity has a clear impact on the FPTD by enhancing longitudinal spreading. This is so because stronger
conductivity heterogeneity leads to broader particle transition time distribution, which in turn leads to enhanced longitudinal
spreading. The injection rule also has a significant impact on FPTDs especially for high conductivity heterogeneity. FPTDs between the two different injection rules are similar for $\sigma_{\ln K} = 1$, but uniform injection shows significantly stronger tailing for $\sigma_{\ln K} = 2, 3, 5$ [Figure~\ref{fig:BTC_MC}(b)]. As conductivity heterogeneity increases, the flux values at the inlet also becomes broader. For flux-weighted injection, most of particles are injected at the nodes with high flux values. However, for uniform injection, particles are uniformly injected across the injection nodes and relatively large number of particles are released at the nodes with low flux values. This leads to notable difference between the two injection rules and the difference grows as the conductivity heterogeneity increases.

\subsection{Centered Mean Square Displacement \label{sec:dispersion}}
\begin{figure}
  \centering
	  \includegraphics[width=5in]{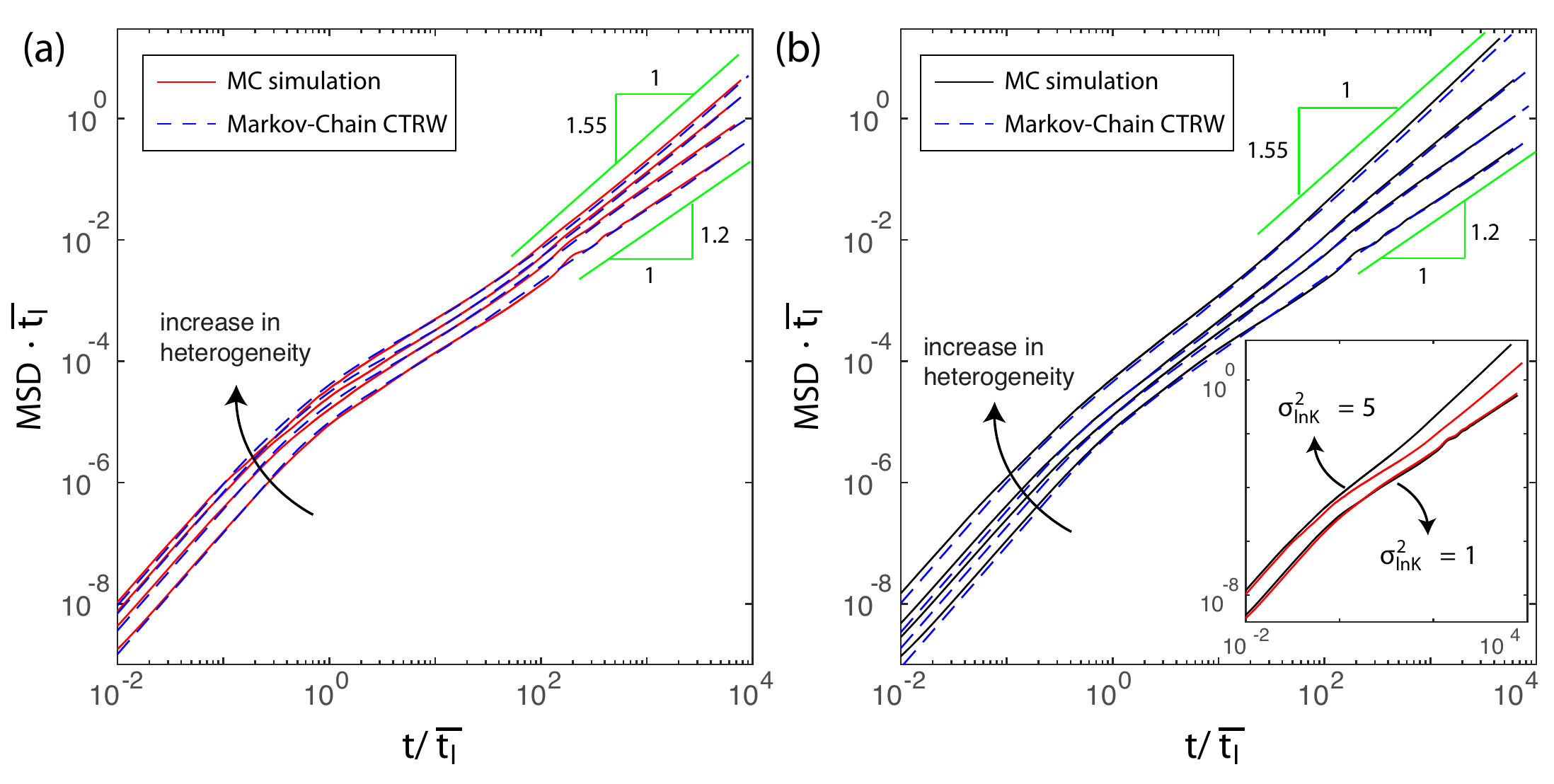}
  \caption{Time evolution of longitudinal MSDs for $\sigma_{\ln K} = 1, 2, 3, 5$ obtained from Monte Carlo simulations (solid lines), and the model predictions from the Markov-Chain CTRW (32) and (35) with the full transition PDF (dashed lines). Increase in conductivity heterogeneity leads to higher dispersion, and the Markov-chain CTRW model is able to accurately capture the time evolution of the MSDs for all levels of heterogeneity and injection rules. (a) Flux-weighted injection. (b) Uniform injection. Inset: Comparison between flux-weighted and uniform injection for $\sigma_{\ln K} = 1, 5$. Impact of injection rule is significant for high conductivity heterogeneity.}\label{fig:moment_prediction}
\end{figure}

We also study longitudinal spreading in terms of the centered mean square
displacement (cMSD) of average particle density, $\overline{ P}(\vx,t)$. For the longitudinal direction ($x$), the cMSD is given by $\sigma^2_{x}(t)=\overline{ \langle[x(t)-\langle x(t)\rangle]^2\rangle}$ where $\langle\cdot\rangle$ denotes the average over all particles for a given realization. In Figure~\ref{fig:moment_prediction}, we show the time evolution of the longitudinal cMSDs.  The time axis is normalized with the mean travel time along the characteristic fracture link length, $\bar{l}$. For both injection methods, spreading shows a ballistic
regime ($\sim t^2$) at early times, which then transitions to a preasymptotic scaling in an intermediate regime and finally to a final asymptotic regime. The time evolutions of cMSDs for the two injection cases are notably different as conductivity heterogeneity increases, while the asymptotic late-time scalings are very similar.  

The asymptotic power-law scaling can be understood in the framework of a continuous time random walk (CTRW) description of dispersion. At large times the Lagrangian velocity distributions are in their steady states and subsequent particle velocities are independent. Thus, at large times horizontal particle dispersion can be described by the CTRW
\begin{align}
\label{ctrw1}
x_{n+1} = x_n + \ell_0, && t_{n+1} = t_n + \tau_n,
\end{align}
with the transition time $\tau_n = {\ell_0} / {v_n}$. The velocities
 $v_n$ are distributed according to $p_s(v)$ which is space Lagrangian velocity PDF, and $\ell_0$ is a distance along the streamline that is sufficiently large so that subsequent particle velocities may be considered independent.   
Thus, the distribution of transit times $\tau_n$ is given in terms of the space Lagrangian and Eulerian velocity PDFs as~\cite[][]{DentzVCTRW2016}
\begin{align}
\label{psise}
\psi(\tau) = \frac{\ell_0}{\tau^2} p_s(\ell_0/\tau) = \frac{\ell_0}{\tau^3 \overline v} p_e(\ell_0/\tau),  
\end{align}
where $\overline v$ is the average Eulerian velocity, see also Section~\ref{Sec:4}. 
Specifically, for the scaling $p_e(v) \propto v^{\alpha}$ at small velocities, the transit time PDF scales as 
\begin{align}
\psi(\tau) \propto \tau^{-1-\beta}, && \beta = 2 + \alpha. 
\end{align}
%
From Figure~\ref{fig:EulerianVpdf}, we estimate for $\sigma_{\ln K} = 5$ that $\alpha \approx -0.55$, which corresponds to $\beta = 1.45$. CTRW theory~\citep{dentzcortis04, berkowitzcortis06} predicts that the cMSD scales as $t^{3-\beta}$, which here implies $t^{1.55}$. This is consistent 
 with the late-time scaling of the cMSD shown in~\ref{fig:moment_prediction} for $\sigma_{\ln K} = 5$. 

The Monte Carlo simulations show that, in the intermediate regime ($t/\overline{t_l}$ approximately between $1$ and  $100$), the \emph{longitudinal} cMSD increases linearly with time for flux-weighted injection [Figure~\ref{fig:moment_prediction}(a)]. For uniform injection, cMSD increases faster than linearly (i.e., superdiffusively) for intermediate to strong heterogeneity in the intermediate regime [Figure~\ref{fig:moment_prediction}(b)]. The stronger heterogeneity led to the increase in the late-time temporal scaling for both flux-weighted and uniform injection cases. The Monte Carlo simulations also show that there is no noticeable difference between the uniform injection and the flux-weighted injection for the low heterogeneity case whereas the difference increases as heterogeneity increases [Figure~\ref{fig:moment_prediction}(b), inset].

In summary, both the increase in conductivity heterogeneity and the uniform injection method enhance longitudinal spreading. For low heterogeneity, the two different injection rules do not affect particle spreading significantly. The difference, however, becomes significant as the conductivity heterogeneity increases. Both the magnitude of the cMSD and the super-diffusive scaling behavior are notably different for the two different injection rules at high heterogeneity. We now analyze the Lagrangian particle statistics to understand the underlying physical mechanisms that lead to the observed anomalous particle spreading.

\section{Lagrangian Velocity Statistics\label{Sec:4}}

The classical CTRW approach~\cite[][]{berkowitzscher97-prl,berkowitzcortis06}---see Eq.~\eqref{ctrw1}---relies on the independence of particle velocities at subsequent steps and thus spatial positions.  Recent studies, however, have shown that the underlying mechanisms of anomalous transport can be quantified through an analysis of the statistics of Lagrangian particle velocities such as velocity distribution and correlation~\citep{leborgnedentz08-prl, meyertchelepi10,dentzbolster10-prl, kangdentz11-prl, deanna13-prl, kangdeanna14, kang15-pre}. 
In the following, we briefly introduce two viewpoints for analyzing Lagrangian velocities---equidistantly and isochronally along streamlines---and the relation between them~\cite[][]{DentzVCTRW2016}. We then proceed to a detailed analysis of the Lagrangian velocity statistics measured equidistantly along streamlines.  

\subsection{Lagrangian Velocities\label{sec:lagrangian}}
Particle motion is described here by the recursion relations~\eqref{eq:langevin}. In this framework,  we consider two types of Lagrangian velocities. The t(ime)--Lagrangian velocities are measured at a given time~$t$,
\begin{align}
v_t(t) = u_{n_t},  
\end{align}
where $n_t$ is defined by~\eqref{nt}. 
The s(pace)--Lagrangian velocities are measured at a given distance $s$ along the trajectory. The distance $s_n$ traveled by a particle along a trajectory after $n$ steps is given by 
\begin{align}
s_{n+1} = s_n + \ell_n. 
\end{align}
The number of steps needed to cover the distance $s$ is described by $n_s = \sup(n|s_n \leq s)$. Thus, the particle velocity at a 
distance $s$ along a trajectory is given by
\begin{align}
v_s(s) = u_{n_s}. 
\end{align}

\begin{figure}
  \centering
	  \includegraphics[width=4in]{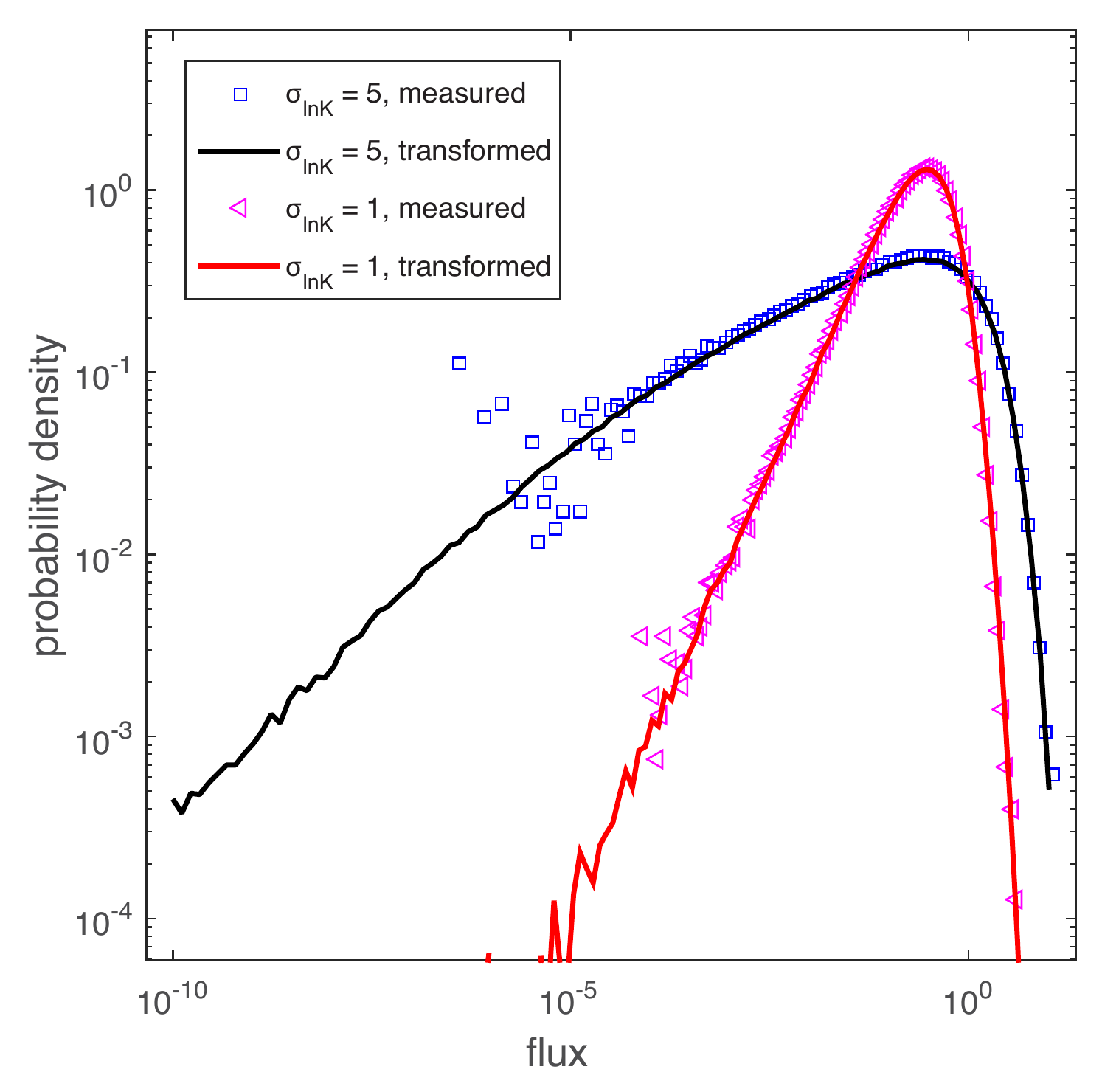}
  \caption{s-Lagrangian velocity PDFs for $\sigma_{\text{ln}K} = 1$ and $\sigma_{\text{ln}K} = 5$. The measured s-Lagrangian velocity PDF agrees very well with the PDF obtained by transforming the Eulerian velocity PDF using Eq.~\eqref{pspe}.}\label{fig:Vpdf_EurToLag}
\end{figure}

The PDF of t--Lagrangian velocities sampled along a particle path is given by
\begin{align}
\label{pt}
p_{t}(v) = \lim_{n \to \infty} \frac{\sum_{i = 1}^n \tau_i \delta(v - u_i)}{\sum_{i=1}^n \tau_i} ,
\end{align}
where we defined the transit time,  
\begin{align}
\label{transit_time}
\tau_i = \frac{\ell_i}{u_i}
\end{align}
The PDF of s--Lagrangian velocities sampled along a particle path
are defined analogously as 
\begin{align}
\label{ps}
p_{s}(v) = \lim_{n \to \infty} \frac{\sum_{i = 1}^n \ell_i \delta(v - u_i)}{\sum_{i=1}^n \ell_i}.
\end{align}
Note the difference with respect to Eq.~\eqref{pe}, which samples velocities in the network uniformly, while in Eq.~\ref{ps} velocities are sampled
along trajectories.  
Using the definition of the transit time $\tau_i$ in~\eqref{transit_time}, the PDFs of the s-- and t--Lagrangian velocities are related through flux weighting as~\cite[][]{DentzVCTRW2016} 
\begin{align}
\label{pspt}
p_s(v) = \frac{v p_t(v)}{\displaystyle \int d v\, v p_t(v)}. 
\end{align}
Furthermore, for flux-preserving flows and under ergodic conditions, the Eulerian and t-Lagrangian velocity PDFs are 
equal,
\begin{align}
p_e(v) = p_t(v).  
\end{align}
Thus, under these conditions, the s--Lagrangian and Eulerian velocity PDFs are related as~\cite[][]{DentzVCTRW2016}  
\begin{align}
\label{pspe}
p_s(v) = \frac{v p_e(v)}{\displaystyle \int d v\, v p_e(v)}. 
\end{align}
This means that the \emph{stationary} s-Lagrangian velocity PDF can be determined from the Eulerian velocity PDF. Figure~\ref{fig:Vpdf_EurToLag} 
illustrates this relation by comparing the s-Lagrangian velocity PDFs measured from the numerical simulation to the flux-weighted Eulerian 
velocity PDFs shown in Figure~\ref{fig:EulerianVpdf}.  

\subsection{Evolution of Lagrangian Velocity Distributions}
It is important to emphasize that the above definitions of the Lagrangian velocity PDFs refer to stationary conditions. 
We now define the PDFs of t-- and s--Lagrangian velocities through sampling between particles and network realizations at a given time (t-Lagrangian) 
or space (s-Lagrangian) velocities
\begin{align}
\hat p_{t}(v,t) = \overline{\langle \delta[v - v(t)] \rangle}, && \hat p_{s}(v,s) = \overline{\langle \delta[v - v(s)] \rangle}. 
\end{align}
In general, these quantities evolve in time and with distance along the streamline and are sensitive to the injection conditions because 
evidently for $t = 0$ and $s = 0$ both are equal to the PDF of initial particle velocities $\hat p_t(v,t=0) = \hat p_s(v,s=0) = p_0(v)$, but their respective 
stationary PDFs are different, namely  
%
\begin{align}
p_t(v) &= \lim_{t \to \infty } \hat p_t(v,t), && p_s(v) = \lim_{s \to \infty} \hat p_s(v,s).  
\end{align}

Let us consider some further consequences of these properties. First, we notice that under (Eulerian) ergodicity the \emph{uniform injection} condition~\eqref{random} corresponds to an initial velocity PDF of 
\begin{align}
p_0(v) = p_e(v) = p_t(v),
\end{align}
that is, the initial velocity PDF is equal to the Eulerian and thus t--Lagrangian velocity PDFs. This means that for the uniform injection method, the t--Lagrangian velocity PDF is steady, $\hat p_t(v,t) = p_t(v)$, while the s--Lagrangian velocity PDF is not. It evolves from its
initial distribution $p_s(v,s=0) = p_e(v)$ to the steady state distribution~\eqref{pspe}. 

The \emph{flux-weighted injection} condition, on the 
other hand, corresponds to the the initial velocity PDF 
\begin{align}
p_0(v) = p_s(v),
\end{align}
due to relation~\eqref{pspt}. The initial velocity PDF is equal to the 
s--Lagrangian velocity PDF. This means that under flux-weighting, the s--velocity PDF is steady, $\hat p_s(v,s) = p_s(v)$. 
Under these conditions, the t--Lagrangian velocity PDF $\hat p_t(v,t)$ evolves from the initial distribution $\hat p_t(v,t=0) = p_s(v)$ towards the asymptotic $p_t(v) = p_e(v)$, which is equal to the Eulerian velocity PDF. These are key insights for the qualitative and quantitative  understanding of the average transport behavior. 

\begin{figure}
  \centering
	  \includegraphics[width=4in]{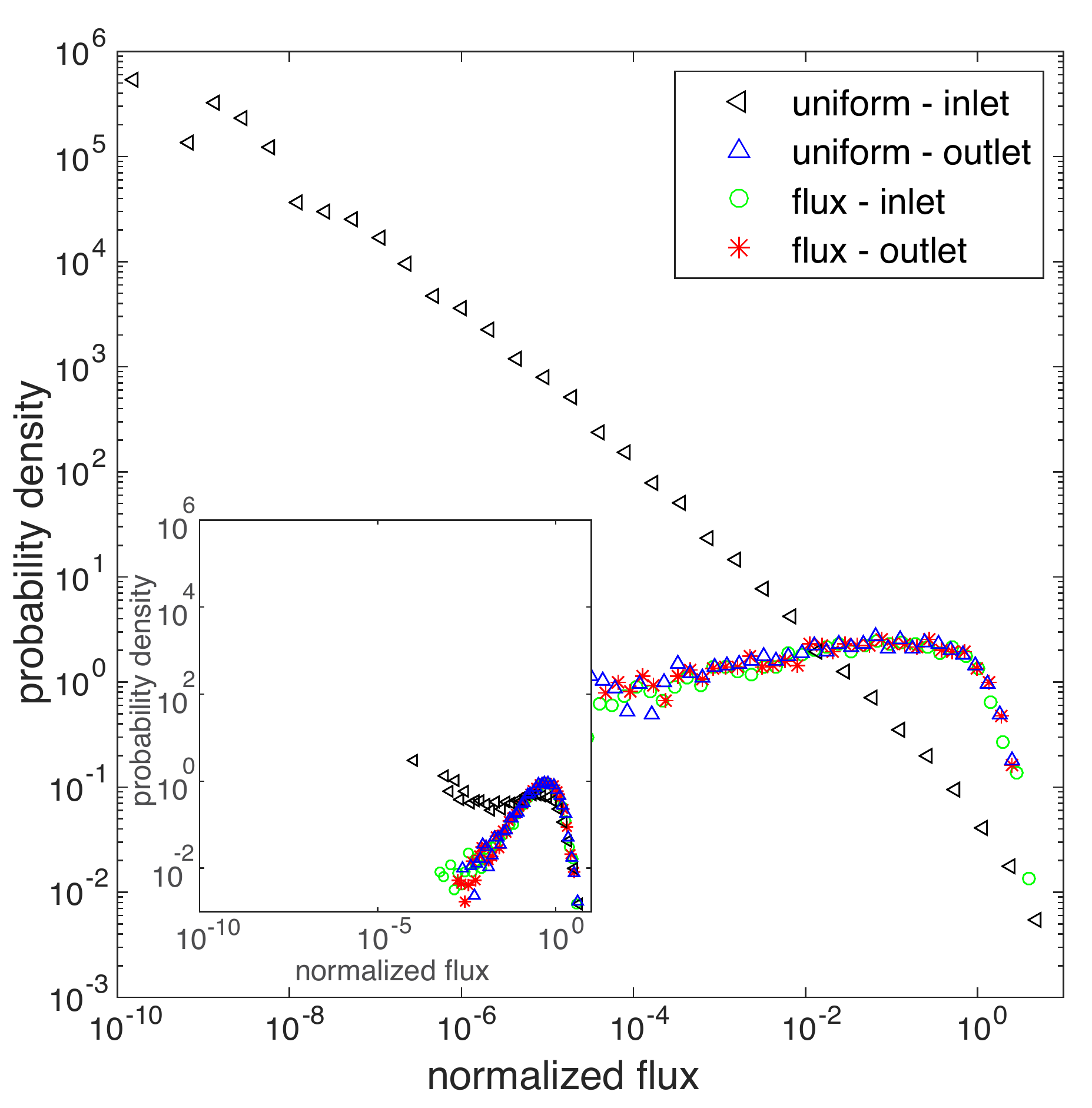}
  \caption{Lagrangian flux distributions at the inlet and outlet for uniform and flux-weighted injection rules, for $\sigma_{\text{ln}K} = 5$. Note that Lagrangian flux distributions at the outlet are identical regardless of the injection method. Inset: the same plot for $\sigma_{\text{ln}K} = 1$. 
  }\label{fig:inletoutlet}
\end{figure}

\subsection{Space-Lagrangian Velocity Statistics}
We analyze particle velocities along their projected trajectories in the longitudinal direction. Spatial particle transitions may be characterized by
 the characteristic fracture link length, $\bar l$. The Lagrangian velocity $v_s(s_n)$ at a 
distance $x_n = n \bar{l}$ along the projected trajectory is approximated by the average velocity 
$v_n \equiv \bar{l}/\tau_n$ where $\tau_n$ is the transition time for the distance $\bar{l}$ at step $n$. In the following, we investigate the statistical characteristics of the s-Lagrangian velocity series $\{v_n\}$. For the uniform flow conditions under consideration here, the projected distance $x_n$ is a measure for the streamwise distance $s_n$, and $v_n$ for the s-Lagrangian velocity $v_s(s_n)$. Spatial Lagrangian velocities have been studied by~\citet{cvetkovic1996} and~\citet{gotovac2009} for highly heterogeneous porous media and by~\citet{frampton2011numerical} for 3D DFNs in view of quantifying particle travel time statistics and thus modelling effective particle motion. 


We first study the convergence of the s--Lagrangian velocity PDFs towards a stationary distribution and the invariance of $\hat p_s(v,s)$ for a stationary (flux-weighted) initial velocity PDF. We consider the two injection conditions~\eqref{random} and~\eqref{flux} and record 
the distribution of particle velocities at a line located at the control point $x_c$.  Under ergodic conditions, we expect $\hat p_s(v,s)$ to converge towards its steady state 
distribution~\eqref{pspe}  for uniform injection and to remain invariant for the flux-weighted injection. Figure~\ref{fig:inletoutlet} shows $\hat p_s(v,s = 0)$ and $\hat p_s(v,s = x_c)$ for uniform and flux-weighted injection conditions and two different heterogeneity strengths. We clearly observe that $\hat p_s(v,x_c) = p_s(v)$ is invariant for flux-weighted injection. For uniform injection, $\hat p_s(v,x_c)$ has already evolved towards its steady limit after $x_c = 200 \bar l$. This is an indication that the flow and transport 
system is in fact ergodic. Note that in terms of computational efficiency, this observation gives a statistically consistent way of continuing 
particle trajectories through reinjection at the inlet. If the outlet is located at a position $x_c$ large enough so that $\hat p_s(v,s = x_c) = p_s(v)$,
particles are reinjected at the inlet with flux-weighted probability, this means that the velocity statistics are preserved. Furthermore, 
this method ensures that the domain is large enough to provide ergodic conditions. In the following, we analyze the statistical properties of streamwise velocity transitions 
with the aim of casting these dynamics in the frame of a Markov model for subsequent particle velocities.    

\begin{figure}
  \centering
	  \includegraphics[width=3.3in]{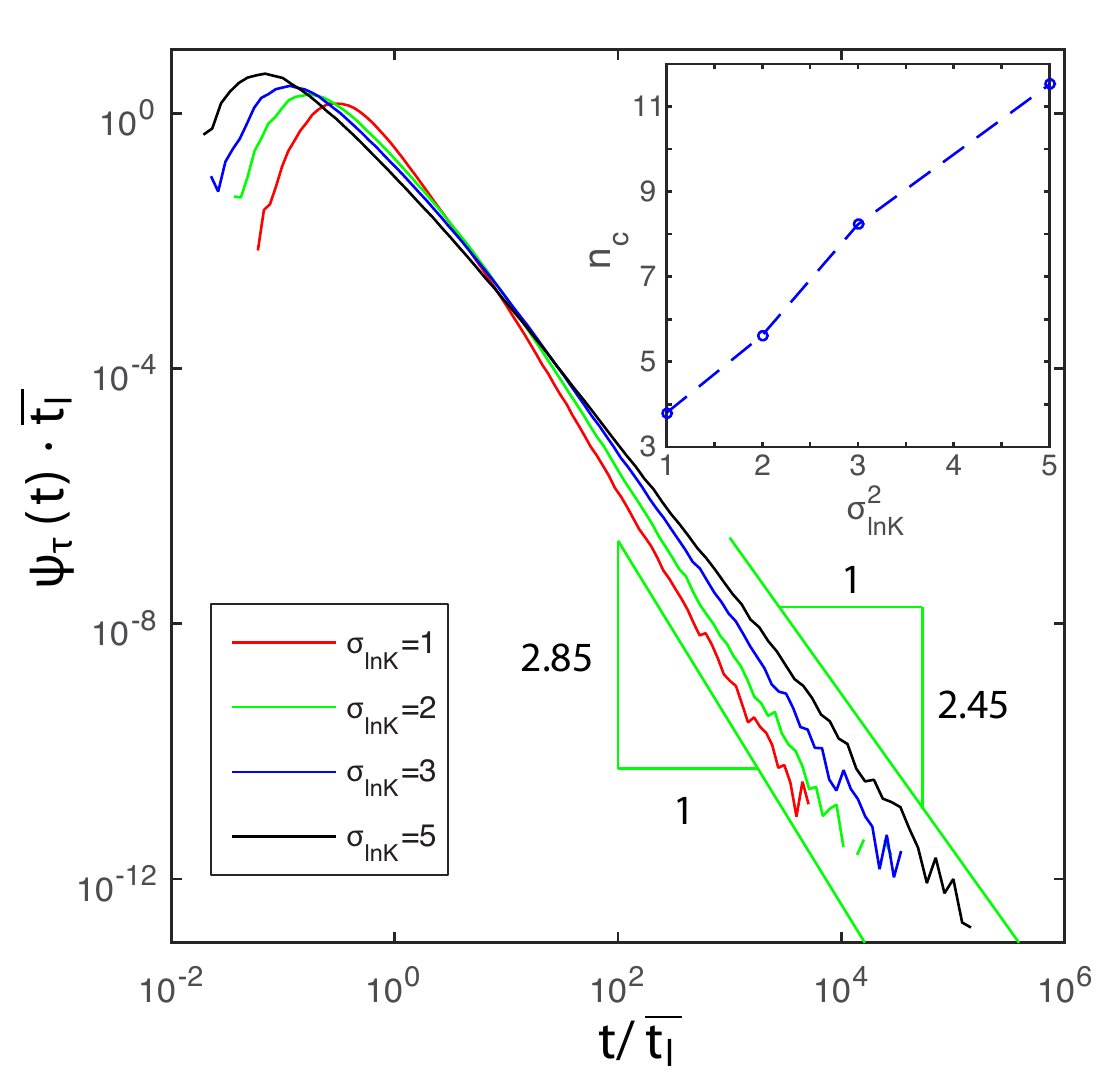}
  \caption{(a) Lagrangian transition time distributions for
   $\sigma_{\ln K} = 1,\;2,\;3,\;5$ with flux-weighted injection. As the network conductivity becomes more heterogeneous, the transition time distribution becomes broader. Inset: the effective correlation length increases with increasing network heterogeneity: $3.8\bar{l}, 5.6\bar{l}, 8.2\bar{l}, 11.5\bar{l}$. The correlation step ($n_c$) is computed by integrating velocity autocorrelation function in space.}
  \label{fig:Tpdf_Vcorr}
\end{figure}

We first consider the distribution $\psi_\tau(t)$ of transition times along particle trajectories through sampling the transition times along all particle trajectories and among network realizations. To this end, we consider a flux-weighted injection because it guarantees that the s-Lagranagian velocities are stationary. Figure~\ref{fig:Tpdf_Vcorr} illustrates the PDF of transition times for different variances of $\ln(K)$. As $\sigma_{\ln K}$ increases, the transition time PDFs become broader. The transition time closely follows a truncated power-law distribution. 

Next we consider two-point velocity statistics to gain insight into the velocity correlations along a streamline. To this end, we consider the velocity auto covariance
for a given lag $\Delta s = s - s^\prime$. As pointed out above, for flux-weighted injection, the streamwise velocities here are stationary and therefore
\begin{align}
C_s(s - s') = \overline{ \langle[v_s(s)-\langle
          v_s(s)\rangle][v_s(s^\prime)-\langle
          v_s(s^\prime)\rangle]\rangle}.
\label{chisp}
\end{align}
In order to increase the statistics, we furthermore sample along streamlines over a distance of $10^2 \bar l$. 
The velocity variance is $\sigma^2_v = C_s(0)$. The velocity autocorrelation function $\chi_s(s) = C_s(s)/\sigma_v^2$.  The correlation length scale $\ell_c$ is defined by
\begin{align}
\ell_c = \int\limits_0^\infty ds\,\chi_s(s). 
\end{align}
The inset in Figure~\ref{fig:Tpdf_Vcorr} shows the increase in the velocity correlation length scale with increasing $\ln(K)$ variances for a flux-weighted injection case. This can be attributed to the emergence of preferential flow paths, as shown in Figure~\ref{fig:flowfield}. \citet{paintercvetkovic05} and \citet{frampton2011numerical} also reported the existence of clear velocity correlation between successive jumps in DFNs and showed that this correlation structure should be captured for effective transport modelling.

The existence of a finite correlation length along the particle trajectories indicates that subsequent velocities, 
when sampled at a distance much larger than  the correlation length $\ell_c$, may be considered independent. In order to study this
feature, we characterize the series of s-Lagrangian velocities $\{v_n\}$ in terms of the transition probabilities to go from velocity $v_m$ to velocity $v_{m+n}$. We determine the transition probabilities under flux-weighted particle injection because, as detailed above, under these conditions, the s-Lagrangian velocity is stationary. Thus, the  transition probability is only a function of the number $n$ of steps,
\begin{align}
r_n(v|v') = \left\langle \delta(v-v_{m+n}) \right\rangle\left|_{v_m = v'}\right. 
\end{align}
Numerically, the transition probability is determined by discretizing the s-Lagrangian velocity PDF into $N$ velocity classes $\mathcal C_i = (v_{s,i},v_{s,i} + \Delta v_{s,i})$ and recording the probability for each class given the previous velocity class. This procedure gives the transition matrix $T_n(i|j)$ from class $j$ to $i$ after $n$ steps such that $r_n(v|v')$ is approximated numerically as
\begin{align}
\label{rT}
r_n(v|v') = \sum_{i,j = 1}^N \frac{\mathbb{I}_{\mathcal C_i}(v) T_{n}(i|j) \mathbb{I}_{\mathcal C_j}(v')}{\Delta v_i},
\end{align}
where the indicator function $\mathbb I_{\mathcal C_i}(v)$ is $1$ if $v \in \mathcal C_i$ and $0$ otherwise. 
\begin{figure}
  \centering
	  \includegraphics[width=4in]{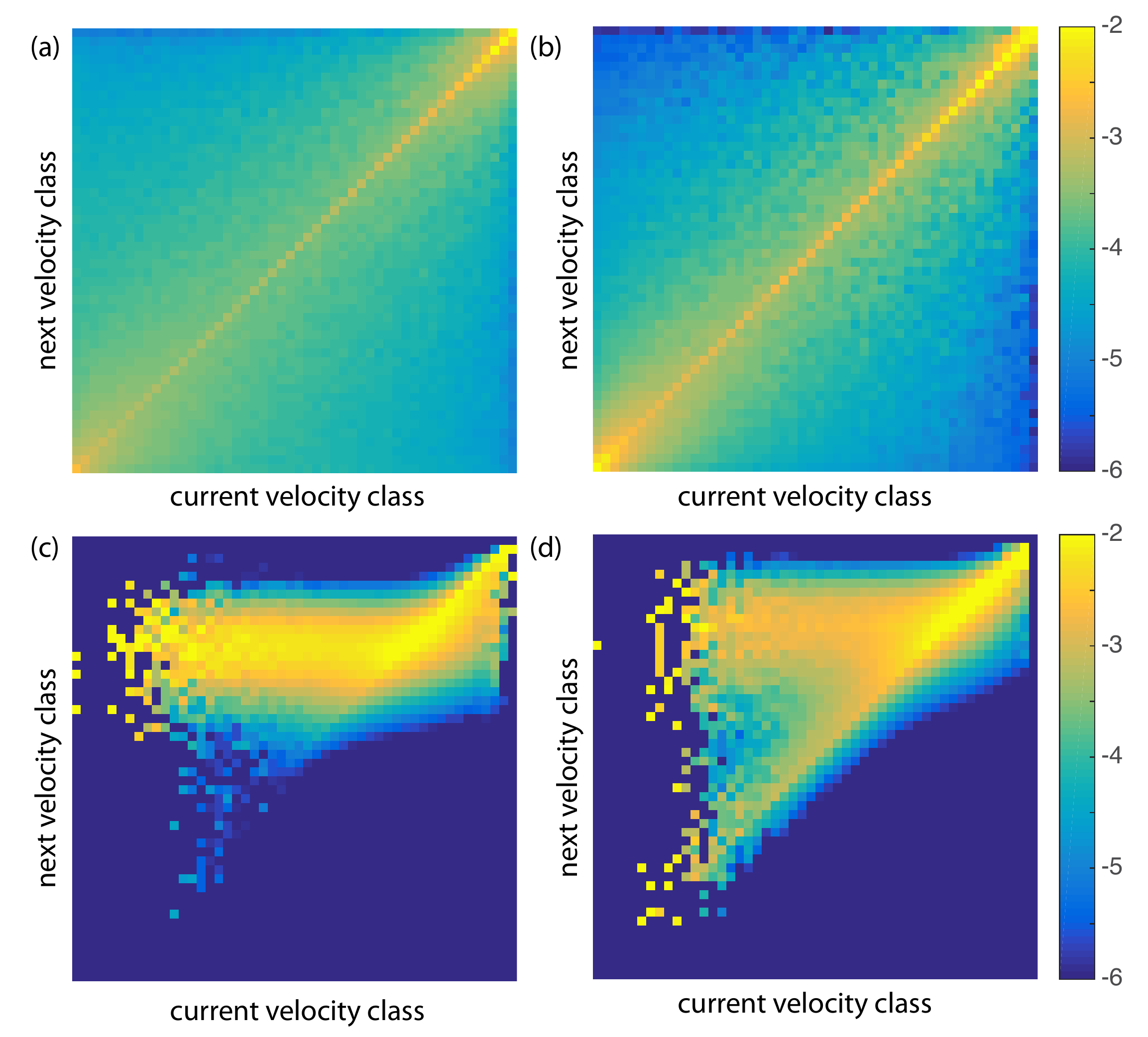}
  \caption{(a)~One-step velocity transition matrix $T_{1}(i|j)$ with linear equiprobable binning for $N=50$ velocity classes for $\sigma_{\ln K} = 1$. (b)~Velocity transition matrix with linear equiprobable binning for $\sigma_{\ln K} = 5$. The color-bar shows the logarithmic scale. (c)~Velocity transition matrix with logarithmic binning for $\sigma_{\ln K} = 1$. (d)~Velocity transition matrix with logarithmic binning for $\sigma_{\ln K} = 5$. Increase in conductivity heterogeneity leads to higher probability close to diagonal entries.}
  \label{fig:Tmat}
\end{figure}

Figure~\ref{fig:Tmat} shows the one-step transition matrix $T_{1}(i|j)$ for equidistant and logarithmically equidistant velocity classes for different network heterogeneity. Higher probabilities along the diagonal than in the off-diagonal positions indicate correlation between subsequent steps, which, however, decreases as the number of steps along the particle trajectory increases, as indicated by the existence of a finite correlation scale $\ell_c$. 


\section{Stochastic Particle Motion and Effective Transport Model\label{Sec:5}}

In the following, we describe the evolution of the s-Lagrangian velocities by a Markov-chain, which is motivated by the existence of a finite spatial correlation scale (see inset of figure \ref{fig:Tpdf_Vcorr}). This leads to a spatial Markov-chain random walk  (which we also termed spatial Markov Model) formulation of particle dispersion that is valid for any initial velocity distribution, and thus for any injection protocol.   
This modelling approach is in line with the time-domain random walk (TDRW) and continuous time random walk (CTRW) approaches discussed in the Introduction and below.


\subsection{Markovian Velocity Process}
\begin{figure}
  \centering
	  \includegraphics[width=4in]{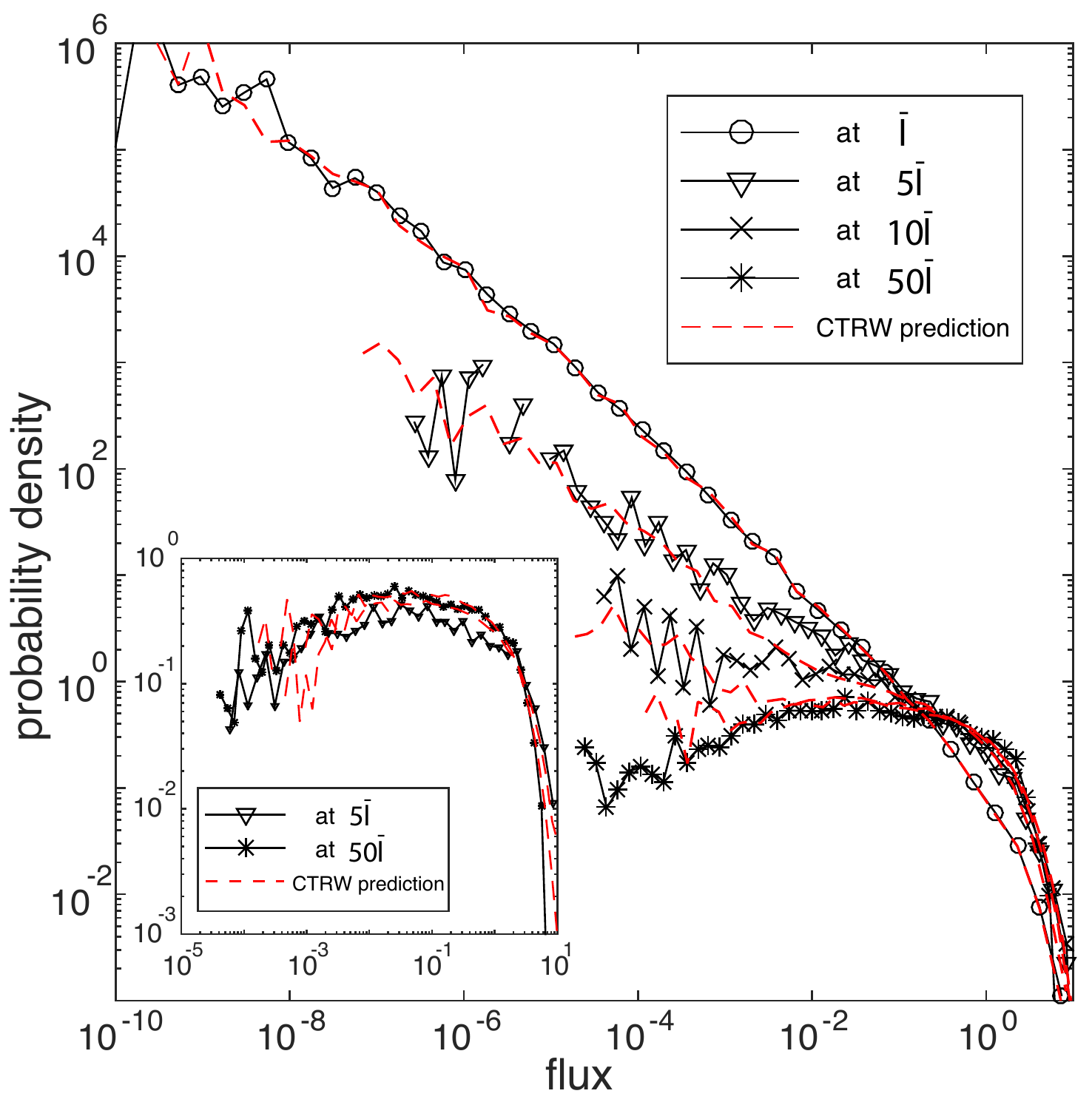}
  \caption{Evolution of the PDF of s-Lagrangian velocities for \emph{uniform injection}, i.e., for an initial velocity PDF $\hat p_s(v,s_0 = 0) = p_e(v)$. The symbols denote the data obtained from the direct numerical simulation, the dashed lines show the predictions of~\eqref{ctrwv} with the transition matrix shown in Figure~\ref{fig:Tmat}. Inset: Evolution of the PDF of s-Lagrangian velocities for a \emph{flux-weighted injection}. In this case, the initial velocity PDF is identical to the stationary s-Lagrangian velocity PDF. \label{fig:vpdf_prediction}}
\end{figure}

Along the lines of~\citet{leborgnedentz08-prl} and~\citet{kangdentz11-prl}, we model the velocity series $\{v_n\}$ as a Markov-chain, which is a suitable model to statistically quantify the evolution of the s-Lagrangian velocities based on the existence of a finite correlation length. In this framework, the $n$--step transition probability $r_n(v|v^\prime)$ satisfies the Chapman--Kolmogorov equation~\cite[][]{Risken:1996}
\begin{subequations}
\label{vMarkovmodel}
\begin{align}
r_{n}(v_n|v_0) = \int d v_k r_{n-k}(v_n|v_k) r_k(v_k|v_0).
\end{align}
The velocity process is fully characterized in terms of the one-step transition PDF $r_1(v|v')$ and the steady state PDF $p_s(v)$ of the s-Lagrangian velocity. Consequently, the evolution of the s-Lagrangian velocity PDF $\hat p_s(v,s_n)$ is given by 
\begin{align}
\label{ctrwv}
p_s(v,s_n) = \int d v' r_{1}(v|v') p_s(v',s_{n}),
\end{align}
\end{subequations}
with the arbitrary initial PDF $p_s(v,s_0 = 0) = p_0(v)$. The number of steps to decorrelate this Markov-chain is given by $n_c = \ell_c/\bar l$. Figure~\ref{fig:vpdf_prediction} shows the evolution of the PDF of s-Lagrangian velocities for the uniform injection~\eqref{random}. Recall that the uniform injection mode corresponds to the initial velocity PDF $p_e(v)$. Thus, the numerical Monte Carlo simulations are compared to the predictions of~\eqref{ctrwv} for the initial condition $\hat p_s(v,s_0 = 0) = p_e(v)$. The transition PDF $r_1(v|v')$ is given by~\eqref{rT} with the velocity transition matrix shown in Figure~\ref{fig:Tmat}. As shown in Figure~\ref{fig:vpdf_prediction}, the prediction of the Markovian velocity model and the Monte Carlo simulation are in excellent agreement, which confirms the validity of the Markov model~\eqref{vMarkovmodel} for the evolution of s-Lagrangian velocities. Velocity transition dynamics are independent of the particular initial conditions and thus allow predicting the evolution of the Lagrangian velocity statistics for any initial velocity PDF and thus for any injection protocol. 

As mentioned above, the Markov-chain $\{v_n\}$ is fully characterized by the stationary PDF of the s-Lagrangian velocities and the transition PDF $r_1(v|v')$. The behavior of the latter may be characterized by the number of steps $n_c$ needed to decorrelate, i.e., the number of steps $n_c$ such that $r_{n}(v|v')$ for $n > n_c$ converges to the stationary PDF $r_n(v|v') \to p_s(s)$. The number of steps for velocities to decorrelate can be quantified by 
\begin{align}
\label{n_c}
n_c = \frac{\ell_c}{\bar l}, 
\end{align}
The simplest transition PDF that shares these characteristics is~\cite[][]{kang15-wrr, kang15-pre, kang2016emergence, DentzVCTRW2016}
\begin{align}
\label{eMarkovmodel}
r_1(v|v') = a \delta(v - v') + (1-a) p_s(v), 
\end{align}
with $ a = \exp(-\bar l/\ell_c)$. This transition PDF is thus fully determined by one single parameter $n_c$. Note that the latter increases with the level of heterogeneity, as illustrated in the inset of figure \ref{fig:Tpdf_Vcorr}. This parameter is estimated here from the simulated Lagrangian velocities. It may also be measured in the field from multiscale tracer tests \citep{kang15-wrr}. 
In the following, we study particle dispersion in the Markovian velocity model for the full transition PDF shown in Figure~\ref{fig:Tmat} and the reduced-order Markov model~\eqref{eMarkovmodel}. 

\subsection{Particle Dispersion and Model Predictions \label{Sec:5.2}}

We consider particle motion along the mean pressure gradient in $x$--direction, which is described by the stochastic regression
\begin{align}
\label{ctrw2}
x_{n+1} = x_n + \bar l, && t_{n+1} = t_n + \frac{\bar l}{v_n}. 
\end{align}
The velocity transitions are determined from the Markovian velocity process~\eqref{vMarkovmodel}. Note that the $\{v_n\}$ process describes equidistant velocity transitions along particle trajectories, while~\eqref{ctrw2} describes particle motion projected on the $x$-axis. In this sense,~\eqref{ctrw2} approximates the longitudinal travel distance $x_n$ with the distance $s_n$ along the streamline, which is valid if the tortuosity of the particle trajectories is low. As indicated in Section~\ref{sec:dispersion}, for travel distances $\ell_0$ larger than $\ell_c$, or equivalently, step numbers $n \gg n_c \equiv \ell_c/\bar l$, subsequent velocities may be considered independent and particle dispersion is fully characterized by the recursion relation~\eqref{ctrw1} and the transition time PDF~\eqref{psise}. Thus, as shown in Section~\ref{sec:dispersion}, the CTRW of Eq.~\eqref{ctrw1} correctly predicts the asymptotic scaling behavior of the centered mean square displacement.  This is not necessarily so for the particle breakthrough and the preasymptotic behavior of the cMSD. As seen in Section~\ref{sec:btc}, the late time tailing of the BTC depends on the injection mode and thus on the initial velocity PDF. 
In fact, the slope observed in Figure~\ref{fig:BTC_MC} for uniform particle injection can be understood through the persistence of the initial velocity PDF. The first random walk steps until decorrelation at $n = n_c$ are characterized by the transit time PDF 
\begin{align}
\psi_0(t) = \frac{\bar l}{t^2} p_0(\bar l/t). 
\end{align}
Thus, for an initial velocity PDF $p_0(v) = p_e(v)$, the initial transit time PDF is given in terms of the Eulerian velocity PDF, which is characterized by a stronger probability weight towards low velocities than the PDF of the s-Lagrangian velocities, which is given by~\eqref{pspe}. The space-time random walk~\eqref{ctrw2} together with the Markov model~\eqref{ctrwv} is very similar to the TDRW approach~\cite[][]{benkepainter03,paintercvetkovic05} and can also be seen as a multi-state, or correlated CTRW approach because subsequent particle velocities and thus transition times are represented by a Markov process~\cite[][]{HausKehr, MonteroPRE2007, leborgnedentz08-prl, ChechkinPRE2009}.  The joint distribution $p(x,v,t)$ of particle position and velocity at a given time $t$ is given by~\cite[][]{leborgnedentz08-prl} 
\begin{align}
\label{p}
p(x,v,t) = \int\limits_0^t d t' H(\bar l/v - t') R(x - v t',v,t-t'), 
\end{align}
where $H(t)$ is the Heaviside step function; $R(x,v,t)$ is the frequency by which a particle arrives at the phase space position $(x,v,t)$. It satisfies 
\begin{align}
\label{R}
R(x,v,t) = R_0(x,v,t) + \int dv' r_1(v|v') R(x - \bar l, v', t- \bar l/v'),
\end{align}
where $R_0(x,v,t) = p_0(x,v) \delta(t)$ with $p_0(x,v) = p(x,v,t = 0)$.  Thus, the right side of~\eqref{p} denotes the probability that a particle arrives at a position $x - v t'$ where it assumes the velocity $v$ by which it advances toward the sampling position $x$. Equations~\eqref{p} and~\eqref{R} can be combined into the Boltzmann equation 
\begin{align}
\frac{\partial p(x,v,t)}{\partial t} = - v \frac{\partial p(x,v,t)}{\partial x}  
- \frac{v}{\bar l} p(x,v,t) + \int dv' \frac{v'}{\bar l} r_1(v|v') p(x, v', t), 
\end{align}
see~\ref{app:boltzmann}. This result provides a bridge between the TDRW approach~\cite{benkepainter03, paintercvetkovic05} and the correlated CTRW approach. 

As illustrated in Figure~\ref{fig:EulerianVpdf}, the Eulerian velocity PDF can be characterized by  the power-law $p_e(v) \propto v^{\alpha}$.  Thus, the first CTRW steps until the decorrelation at $n = n_c$ are characterized by the transit time PDF $\psi_0(t) \propto t^{-2 - \alpha}$. The corresponding tail of the BTC is $f(t,x_c) \propto t^{-2-\alpha}$.  The observed value of $\alpha = -0.55$ explains the tailing of the BTC in Figure~\ref{fig:BTC_MC} as $t^{-1.45}$, which shows the importance of the initial velocity distribution. We also observe decrease in BTC tailing (larger absolute slope) with travel distance as initial velocity distribution converges to stationary Lagrangian velocity distribution and as tracers sample more velocity values. This implies that one needs to be careful when inferring a $\beta$ from single BTC measurement because the slope can evolve depending on the injection method, velocity PDF and velocity correlation.

\begin{figure}
  \centering
	  \includegraphics[width=5in]{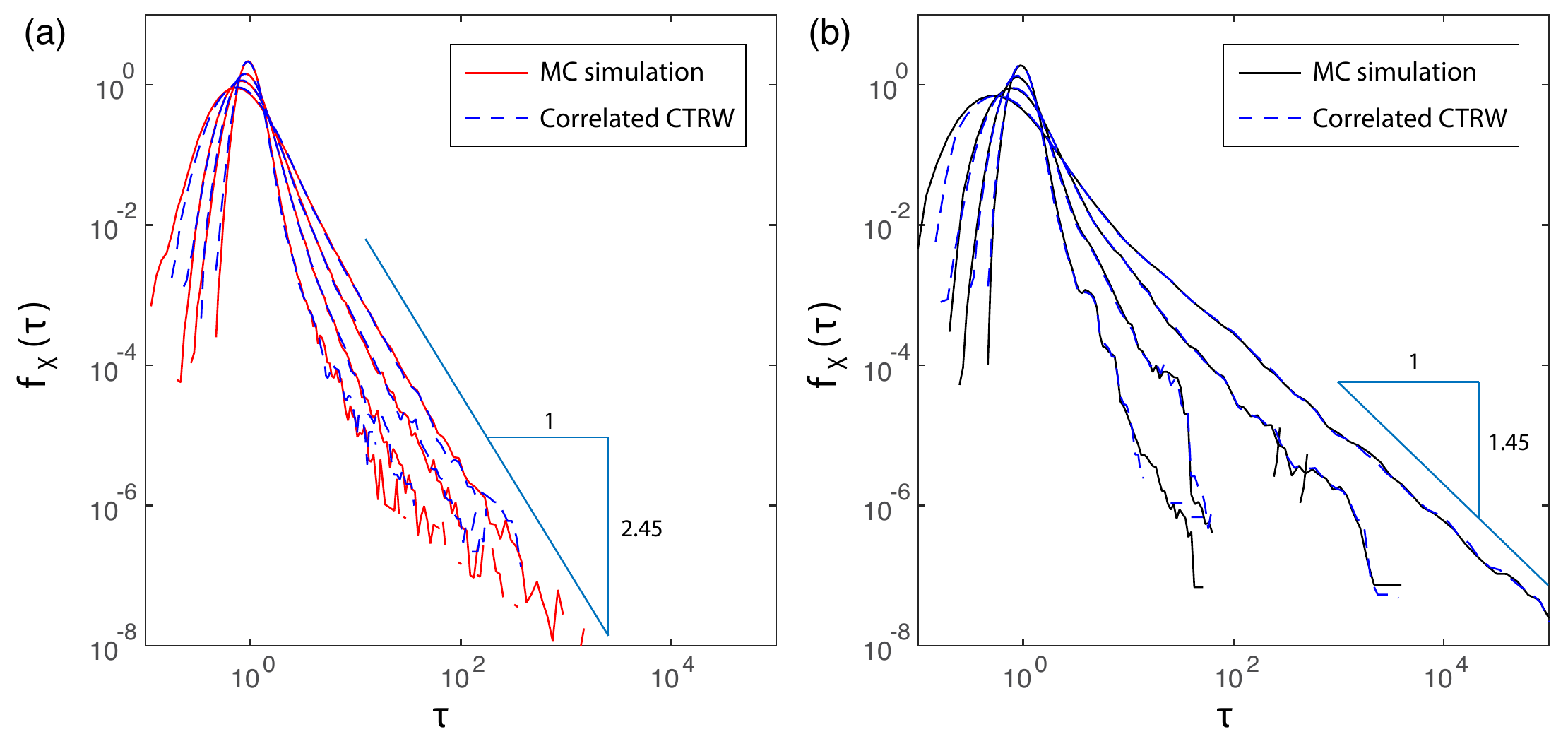}
  \caption{Particle BTCs from
    Monte Carlo simulations and the predictions from the Markov-chain CTRW model with the \emph{full} velocity transition matrix for (a)~flux-weighted
    injection, and (b)~uniform injection at $x_c = 200 \bar l$.}\label{fig:BTC_prediction}
\end{figure}

First, we compare the results obtained from Monte Carlo simulation in the random DFN to the predictions of the Markov-chain CTRW~\eqref{vMarkovmodel} and \eqref{ctrw2} with the full transition PDF of Figure~\ref{fig:Tmat}.
Figure~\ref{fig:moment_prediction} shows the evolution of the cMSD for different levels of heterogeneity and different injection modes. As expected from the ability of the Markov model to reproduce the evolution of the s-Lagrangian velocity PDF for both uniform and flux-weighted injection conditions, the predictions of particle spreading are in excellent agreement with the direct numerical simulations. In Figure~\ref{fig:BTC_prediction} we compare breakthrough curves obtained from numerical simulations with the predictions by the Markov model for the uniform and flux-weighted injection modes. Again, the impact of the injection mode and thus initial velocity PDF is fully quantified by the Markov model. 

\begin{figure}
  \centering
	  \includegraphics[width=5in]{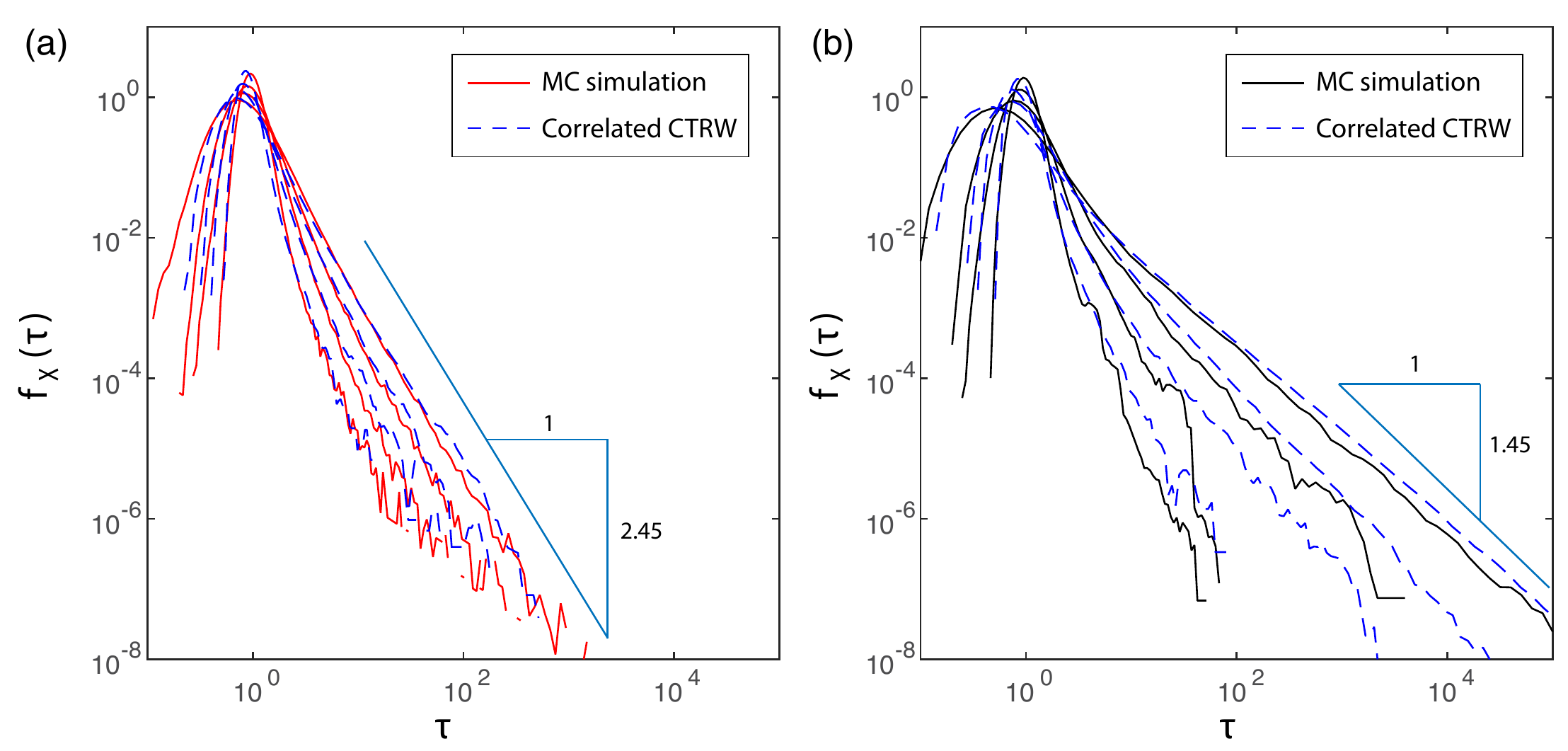}
  \caption{Particle BTCs from
    Monte Carlo simulations and predictions from the Markov-chain CTRW model with the \emph{reduced-order} velocity transition matrix for (a)~flux-weighted
    injection, and (b)~uniform injection at $x_c = 200 \bar l$.}\label{fig:BTC_prediction_eff}
\end{figure}

We now apply the Markov model~\eqref{vMarkovmodel}--\eqref{ctrw2}, i.e., employing a parsimonious parameterization of the velocity transition PDF, with a single parameter $n_c$ (equation \eqref{n_c}), which is estimated here from velocity correlations along streamlines (see inset of figure \ref{fig:Tpdf_Vcorr}). 
We first compare the reduced-order Markov model to the cases of uniform and flux-weighted injection, and conclude that the proposed parsimonious stochastic model provides an excellent agreement with the direct numerical simulations  (Figure~\ref{fig:BTC_prediction_eff}). This implies that the simple correlation model~\eqref{eMarkovmodel} can successfully approximate the velocity correlation structure. Hence it appears that high order correlation properties, quantified from the full transition probabilities (figure \ref{fig:Tmat}), are not needed for accurate transport predictions in the present case.  This suggests promising perspective for deriving approximate analytical solutions for this Markov-chain CTRW model \cite{DentzVCTRW2016}. Furthermore, as discussed in \cite{kang15-wrr}, the velocity correlation parameter $n_c$ can be estimated in the field by combining cross-borehole and push-pull tracer experiments.

Finally, we consider the evolution of the particle BTC and the cMSD for \emph{arbitrary injection modes}. For real systems both flux-weighted and uniform injections are idealizations. A flux-weighted condition simulates a constant concentration pulse where the injected mass is proportional to the local injection flux at an inlet boundary that is extended over a distance much larger than the correlation scale during a given period of time. A uniform injection represents an initial concentration distribution that is uniformly extended over
a region far larger than the correlation length. In general, the initial concentration distribution may not be uniform, and the injection boundary may not be sufficiently large, which leads to an arbitrary initial velocity distribution, biased maybe to low or high flux zones, as for example in the MADE experiments,
where the solute injection occurred into a low permeability zone~\citep{harvey2000}. For demonstration, we study two scenarios representing injection into low and high flux zones: uniform injections into regions of the 20-percentile highest, and 20-percentile lowest velocities. The initial velocity PDF for the low velocity mode shows the power-law behavior which is the characteristic for the Eulerian PDF, and the initial velocity PDF for the high velocity mode shows narrow initial velocity distribution (Figure 15). Eventually, the s-Lagrangian PDFs evolve towards the stationary flux-weighted Eulerian PDF as discussed in the previous section.
 
Figure~\ref{fig:BTC_MSD_topbottom_predict} shows the predictive ability of the effective stochastic model for these different injection conditions. The reduced-order Markov velocity model compares well with the direct Monte Carlo simulation in the random networks. As expected, the BTCs for injection into low velocity regions have a much stronger tailing than for injection into high velocity regions. In fact, as the initial velocity shows the same behavior at low velocities as the Eulerian velocity PDF, the breakthrough tailing is the same as observed in Figure~\eqref{fig:BTC_prediction}. We also observed that the reduced-order Markov velocity model can capture important features of the time evolution of cMSDs. This demonstrates that the proposed model can incorporate arbitrary injection modes into the effective modelling framework.

\begin{figure}
  \centering
	  \includegraphics[width=4in]{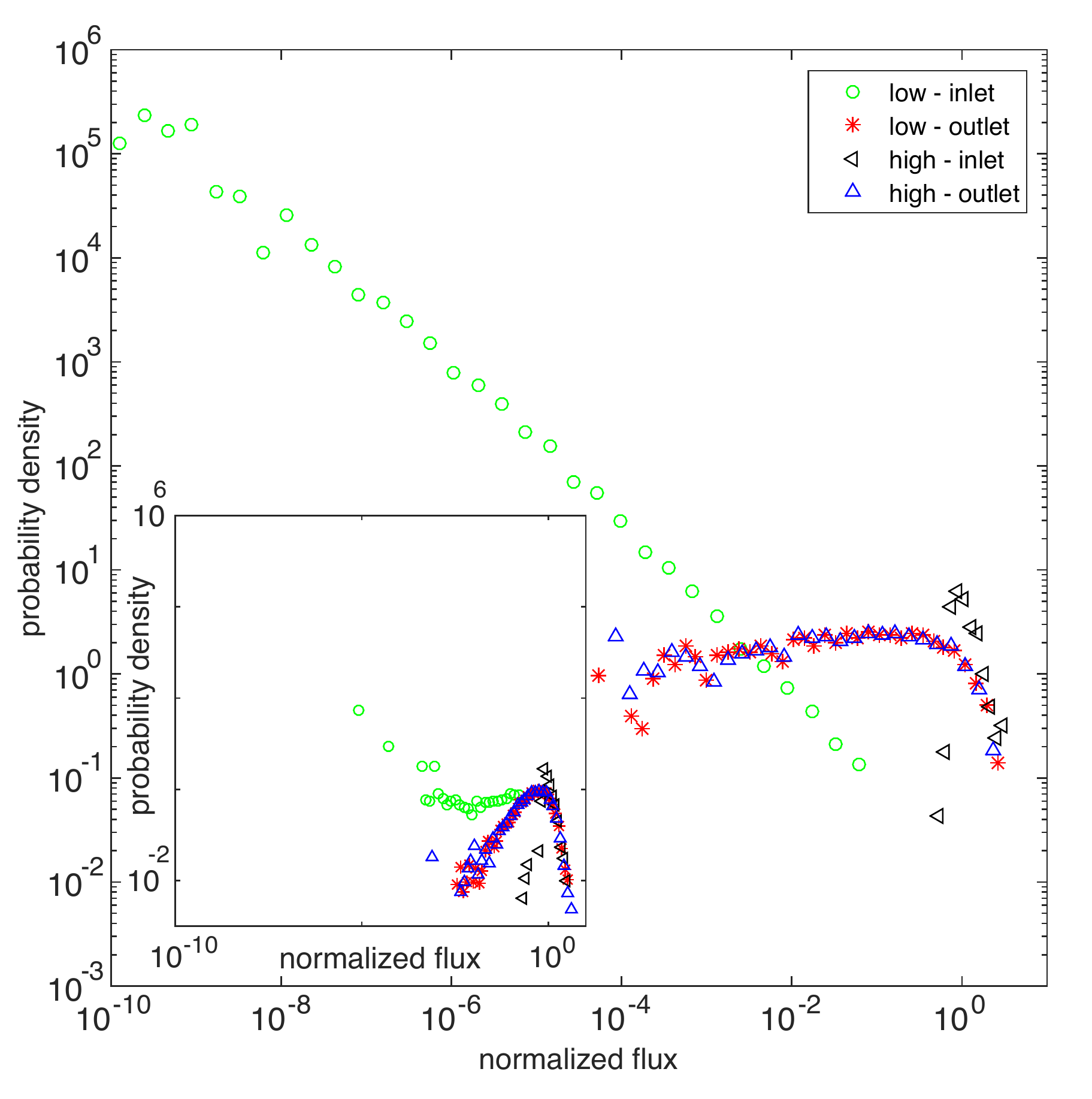}
  \caption{Lagrangian flux distributions at the inlet and outlet with two arbitrary initial velocity distributions for $\sigma_{\ln K} = 5$. The two initial velocity distributions come from uniform injections into regions of the 20-percentile highest, and 20-percentile lowest velocities. Flux values are normalized with the mean flux value. Note that flux distributions at outlet are identical regardless of the initial velocity distribution. Inset: same plot for $\sigma_{\ln K} =1$.}\label{fig:vhighlow}
\end{figure}

\begin{figure}
  \centering
	  \includegraphics[width=5in]{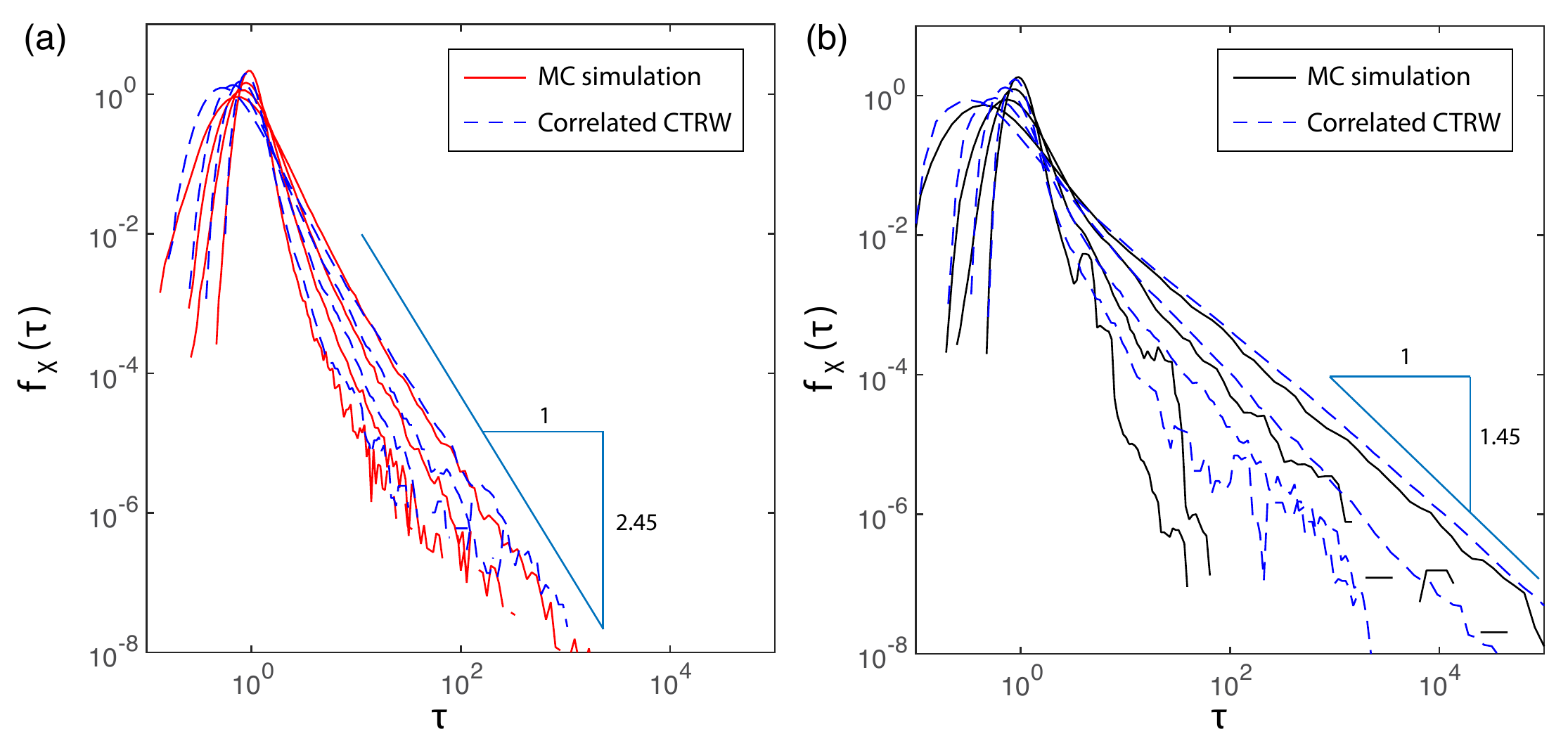}
  \caption{(a)~Particle BTCs from Monte Carlo simulations for injection into \emph{high-velocity} regions (solid line) and predictions from Markov-chain CTRW model with the reduced-order velocity transition matrix (dashed line). (b)~Corresponding results for injection into \emph{low-velocity} regions.}\label{fig:BTC_MSD_topbottom_predict}
\end{figure}

\section{Conclusions\label{Sec:6}}
This study shows how the interplay between fracture geometrical properties (conductivity distribution and network geometry) and tracer injection modes controls average particle transport via Lagrangian velocity statistics. The interplay between fracture heterogeneity and tracer injection methods can lead to distinctive anomalous transport behavior. Furthermore, the injection conditions, for example, 
uniform or flux-weighted, imply different initial velocity distributions, which can have a persistent impact on particle spreading through DFNs. For uniform injection, the s-Lagrangian velocity distribution evolves from an Eulerian velocity distribution initially to a stationary s-Lagrangian distribution. In contrast, for flux-weighted injection, the s-Lagrangian velocity distribution remains stationary.  

We have presented a spatial Markov model to quantify anomalous transport through DFNs under arbitrary injection conditions. We derive an analytical relation between the stationary Lagrangian and the Eulerian velocity distribution, and formally incorporate the initial velocity distribution into the spatial Markov model. The proposed model accurately reproduces the evolution of the Lagrangian velocity distribution for arbitrary injection modes. This is accomplished with a reduced-order stochastic relaxation model that captures the velocity transition with a single parameter: the effective velocity correlation~$\ell_c$. 
The agreement between model predictions and direct numerical simulations indicates that the simple velocity correlation model can capture the dominant velocity correlation structure in DFNs.

In this study, we investigated the particle transport and the impact of the injection condition for idealized 2D DFN using a Markov velocity model. These findings can be extended to 3D DFNs, for which similar behaviors regarding the injection mode have been found~\cite{hyman2015influence}. Also,~\citet{frampton2011numerical} reported similar velocity correlation structures for 3D DFNs as in 2D, which suggests that a velocity Markov model such as the one presented in this work can be used for the modelling of particle motion in 3D DFNs. 
 

\bigskip
{\bf Acknowledgements:} PKK and SL acknowledge a grant (16AWMP-B066761-04) from the AWMP Program funded by the Ministry of Land, Infrastructure and Transport of the Korean government and the support from Future Research Program (2E27030) funded by the Korea Institute of Science and Technology (KIST). PKK and RJ acknowledge a MISTI Global Seed Funds award. MD acknowledges the support of the European Research Council (ERC) through the project MHetScale (617511). TLB acknowledges the support of European Research Council (ERC) through the project ReactiveFronts (648377). RJ acknowledges the support of the US Department of Energy through a DOE Early Career Award (grant DE-SC0009286). The data to reproduce the work can be obtained from the corresponding author.

\appendix
\section{Boltzmann Equation\label{app:boltzmann}}
The time derivative of~\eqref{p} gives 
\begin{align}
\label{papp}
\frac{\partial p(x,v,t)}{\partial t} = - v \frac{\partial p(x,v,t)}{\partial x} + R(x,v,t) - R(x - \bar l,v,t - \bar l/v). 
\end{align}
Note that $R(x,v,t)$ denotes the probability per time that a particle has the velocity $v$ at the position $x$. 
It varies on a time scale of $\bar \/v$. Thus we can approximate~\eqref{p} as 
\begin{align}
p(x,v,t) \approx \frac{\bar l}{v} R(x - \bar l,v,t-\bar l/v). 
\end{align}
Using this approximation and combining~\eqref{papp} with~\eqref{R} gives for $t > 0$
\begin{align}
\frac{\partial p(x,v,t)}{\partial t} = - v \frac{\partial p(x,v,t)}{\partial x}  
- \frac{v}{\bar l} p(x,v,t) + \int dv' r_1(v|v') \frac{v'}{\bar l} p(x, v', t). 
\end{align}
%
\bibliographystyle{elsarticle-num-names}

\end{document}